\documentclass[aps,prd,amsmath,amsfonts,amssymb,eqsecnum,nofootinbib,
  floatfix,secnumarabic,preprintnumbers,superscriptaddress,nobalancelastpage,
  onecolumn,notitlepage]{revtex4-1}
\usepackage[utf8]{inputenc}
\usepackage{color,graphicx,hyperref,dsfont,slashed,multirow}

\hypersetup{bookmarks=true,unicode=true,pdftoolbar=true,pdfmenubar=true,
  pdffitwindow=false,pdfstartview={FitH},pdftitle={HeavyHiggs},
  pdfauthor={B.~Coleppa,~B.~Fuks,~P.Poulose,~S.~Sahoo},
  pdfsubject={Subject},pdfcreator={Creator},pdfproducer={Producer},
  pdfkeywords={Heavy Higgs bosons}{2HDM},pdfnewwindow=true,colorlinks=true,
  linkcolor=blue,citecolor=magenta,filecolor=magenta,urlcolor=cyan}
\def\bsp#1\esp{\begin{split}#1\end{split}}
\def\be{\begin{equation}}
\def\ee{\end{equation}}
\def\bpm{\begin{pmatrix}}
\def\epm{\end{pmatrix}}

\def\h{h^0}
\def\H{H^0}
\def\A{A}
\newcommand{\sba}{\ensuremath{\sin(\beta-\alpha)}}
\newcommand{\cba}{\ensuremath{\cos(\beta-\alpha)}}
\newcommand{\hc}{H^{\pm}}

\begin{document}
\date{\today}

\title{Seeking Heavy Higgs Bosons through Cascade Decays}

\author{Baradhwaj Coleppa}
\email {baradhwaj@iitgn.ac.in}
\affiliation {Physics Discipline,
Indian Institute of Technology-Gandhinagar, Palaj Campus, Gujarat 382355, India}

\author{Benjamin Fuks}
\email{fuks@lpthe.jussieu.fr}
\affiliation{Sorbonne Universit\'es, UPMC Univ.~Paris 06, UMR 7589, LPTHE,
  F-75005, Paris, France}
\affiliation{CNRS, UMR 7589, LPTHE, F-75005, Paris, France}
\affiliation{Institut Universitaire de France, 103 boulevard Saint-Michel,
  75005 Paris, France}

\author{P. Poulose}
\email {poulose@iitg.ernet.in}
\author{Shibananda Sahoo}
\email {shibananda@iitg.ernet.in}
\affiliation {Department of Physics,
Indian Institute of Technology-Guwahati, Assam 781039, India}

\begin{abstract}
  We investigate the LHC discovery prospects for a heavy Higgs boson decaying
  into the Standard Model Higgs boson and additional weak bosons. We consider a
  generic model-independent new physics configuration where this decay proceeds
  via a cascade involving other intermediate scalar bosons and focus on an LHC
  final-state signature comprised either of four $b$-jets and two charged
  leptons or of four charged leptons and two $b$-jets. We design two analyses of
  the corresponding signals, and demonstrate that a $5\sigma$ discovery at the
  14~TeV LHC is possible for various combinations of the parent and daughter
  Higgs-boson masses. We moreover find that the Standard Model backgrounds can
  be sufficiently rejected to guarantee the reconstruction of the parent Higgs
  boson mass. We apply our analyses to the Type-II Two-Higgs-Doublet Model and
  identify the regions of the parameter space to which the LHC is sensitive.
\end{abstract}

\maketitle

\section{Introduction}
The discovery of a Higgs boson whose properties are consistent with the
expectations of the Standard Model (SM) has undoubtedly been the triumph of the
Large Hadron Collider (LHC) thus far~\cite{Aad:2012tfa,Chatrchyan:2012xdj,%
Khachatryan:2016vau,Aad:2015zhl}. While it is clear that the Higgs boson plays a
central role in the breaking of the electroweak symmetry, there is still room
for a non-minimal Higgs sector with a more involved TeV scale structure than
what could be expected from the SM alone. Moreover, the proof of
existence of the Higgs boson has provided an additional tool to narrow down the
possibilities for new physics, additional constraints on the new physics
parameter spaces being imposed by enforcing the Higgs-boson branching ratio and
production cross section predictions to agree with the measured values. However,
the Higgs boson could also be a perfect laboratory for uncovering new
physics in cases where new heavier particles could decay into it, enhancing its
indirect production rate.

One attractive minimal scenario along these lines is the so-called
Two-Higgs-Doublet Model (2HDM) where the SM Higgs sector is extended by a second
weak doublet of Higgs fields~\cite{Lee:1973iz,Branco:2011iw}, although numerous
not so minimal options like the Minimal Supersymmetric Standard Model~\cite{%
Nilles:1983ge,Haber:1984rc} or the Next-to-Minimal Supersymmetric Standard
Model~\cite{Ellwanger:2009dp} fall into that category of models as well.
The physical spectrum then (minimally) contains, on top of the Standard
Model Higgs boson $\h$, a heavy scalar field $\H$, a pseudoscalar field $\A$ as
well as a pair of charged Higgs bosons $\hc$. A general feature of heavier
Higgs bosons consists in the dominance of Higgs-to-Higgs decays in association
with a weak boson as soon as they are kinematically open~\cite{Moretti:2016jkp,%
Arhrib:2016wpw,Coleppa:2014cca}. This has consequently motivated the search for
the corresponding signals in LHC data by both the ATLAS~\cite{Aad:2015wra,%
TheATLAScollaboration:2016loc,ATLAS:2017nxi,Aaboud:2017ahz} and CMS~\cite{%
Khachatryan:2015lba,Khachatryan:2016are,CMS:2016qxc} collaborations, as well as
a series of theoretical works both in the 2HDM~\cite{Yang:2011jk,Chen:2014dma,%
Coleppa:2014hxa,Dumont:2014wha,Bernon:2015wef,Bernon:2015qea,Kling:2015uba,%
Li:2015lra,Li:2016umm,Patrick:2016rtw,Goncalves:2016qhh,%
Kling:2016opi,Chakrabarty:2017qkh,Ferreira:2017bnx} and other (less minimal) new
physics models~\cite{Kang:2013rj,Hajer:2015gka,Casolino:2015cza,Conte:2016zjp,%
Aggleton:2016tdd,Ellwanger:2017skc}.

In this work, we generalize this concept of Higgs-to-Higgs decays when several
weak bosons arise from the cascade~\cite{Craig:2012pu,Gao:2016ued}, like when
in the 2HDM, the heaviest scalar Higgs boson $\H$ decays via a lighter
pseudoscalar state $\A$ into a SM Higgs boson, $\H\to\A Z\to \h Z Z$. We further
consider SM boson decays into leptons or jets
originating from the fragmentation of $b$-quarks, as the latter consists of the
dominant decay mode of the SM Higgs boson. In particular, we focus on a
final-state signature made of either four leptons and two b-jets, or of two
leptons and four $b$-jets, and we analyze the corresponding LHC prospects. We
first consider a simplified model approach (Section~\ref{sec:indep}) that can
easily be reinterpreted into numerous models featuring an extended Higgs sector.
In Section~\ref{sec:model}, we take the example of the Type-II 2HDM and
translate
our findings in the corresponding parameter space. We summarize our work and
present our conclusions in Section~\ref{sec:conclusions}.

\section{A Simplified Model for Analysing Higgs Cascade Decay Signals}
\label{sec:indep}

\subsection{Theoretical Framework, Benchmark Scenarios and Simulation Setup}
In order to determine the LHC sensitivity to Higgs-to-Higgs cascade decays, we
make use of a simplified model where the SM is minimally extended in
terms of new particles and couplings. In practice, we complement the SM field
content by two additional scalar bosons, so that the scalar part of the particle
spectrum now contains the observed SM-like Higgs boson $\h$ and two new
states that we denote by $H_1$ and $H_2$. In our convention, $H_1$ is the
lighter boson and $H_2$ the heavier one, and the couplings of the new scalars to
the $Z$-boson are kept generic. Whilst their strengths are in principle free
parameters, they are traded, in the analyses of the next subsections, for the
signal cross sections (see below for more details). We assume varied mass
differences between the two new states and the SM Higgs boson, so that we define
four different scenarios that we name
{\bf BP1}, {\bf BP2}, {\bf BP3} and {\bf BP4} and for which the heavy scalar
masses $m_{H_1}$ and $m_{H_2}$ read
\be\bsp
  {\rm \bf BP1}:\quad m_{H_1} = 250~{\rm GeV},\qquad&m_{H_2} = 400 ~{\rm GeV},\\
  {\rm \bf BP2}:\quad m_{H_1} = 600~{\rm GeV},\qquad&m_{H_2} = 1000~{\rm GeV},\\
  {\rm \bf BP3}:\quad m_{H_1} = 250~{\rm GeV},\qquad&m_{H_2} = 1000~{\rm GeV},\\
  {\rm \bf BP4}:\quad m_{H_1} = 400~{\rm GeV},\qquad&m_{H_2} = 600 ~{\rm GeV}.
\esp\ee
This choice of benchmark points allows us to capture various features that
could arise from distinct
mass-splitting options. In the {\bf BP1} scenario, there is not much available
phase space for both the $H_2\rightarrow ZH_1$ and $H_1\rightarrow \h Z$ decays
and thus these occur close to threshold. In contrast, the larger mass splittings
featured by
the {\bf BP2} configuration, in which $m_{H_2} \gg m_{H_1} \gg m_\h$, implies
that both the $H_1$ and $H_2$ decays proceed far from threshold, the decay
products being thus expected to feature a larger amount of transverse momentum
$p_T$. The third scenario {\bf BP3} consists of an intermediate case where
only the  $H_1\rightarrow Z\h$ decay occurs close to threshold. Finally, in
the fourth scenario {\bf BP4}, both decays occur far from threshold, but the
mass splitting is reduced compared to the {\bf BP2} case.

The different mass splittings between the $\h$, $H_1$ and $H_2$ states probed in
our benchmarks are expected to impact the kinematic properties of the leptons
and $b$-jets originating from the decays of the final-state SM Higgs boson and
$Z$-bosons. As a consequence, their study could provide handles for unraveling
new physics at the LHC. In the following, we consider the production of the
heaviest Higgs boson $H_2$ through gluon fusion, and its subsequent decays into
lighter Higgs states and $Z$-bosons,
\be
 p p \to H_2 \to H_1 Z \to \h Z Z \ .
\label{eq:process}\ee
Whilst we focus on the dominant Higgs boson decay mode $\h\to b\bar b$, we
consider $Z$-boson decays into a lepton pair $Z\to\ell^+ \ell^-$ and into a
bottom-antibottom pair $Z\to b\bar b$. Omitting a final-state signature
comprised of six $b$-jets, given the huge associated multijet background and the
difficulties induced by the combinatorics to reconstruct all intermediate
particles, the final-state signatures of interest therefore consist of a system
made either of four leptons and two $b$-jets ($4\ell 2b$) or of two leptons and
four $b$-jets ($2\ell 4b$).

As above-mentioned, the signal cross section is taken as a free parameter which
correspondingly allows us to ignore the actual strengths of the $Z$-boson
couplings to the new scalar bosons. As a benchmark, we make use of a fiducial
signal cross sections $\sigma_{\rm fid}$ fixed to
\be
  \sigma_{\rm fid}(p p \to H_2 \to H_1 Z \to \h Z Z \to 4\ell 2b) = 5~{\rm fb}
  \quad\text{and}\quad
  \sigma_{\rm fid}(p p \to H_2 \to H_1 Z \to \h Z Z \to 2\ell 4b) = 5~{\rm fb}
  \ ,
\ee
whch consist of values lying in the ball park of what could be achieved in a
phenomenologically-viable model.

Hard-scattering signal events at a collision center-of-mass energy of 14~TeV are
generated by means of the {\sc MadGraph}5\_{aMC@NLO} platform~\cite{%
Alwall:2014hca}. Practically, we convolute the signal leading-order matrix
elements, as
automatically obtained from the 2HDM UFO~\cite{Degrande:2011ua} model available
from the {\sc FeynRules} repository~\cite{Christensen:2009jx,Alloul:2013bka},
with the leading-order set of NNPDF parton densities version 3.0~\cite{%
Ball:2014uwa}.
The dependence on the numerical values of the different coupling strengths being
factorized out by an appropriate choice of the fiducial cross section, the sole
model dependence consists of the Lorentz structure of the various interactions
of the $Z$-boson with Higgs bosons. This restriction is however compatible with
a large variety of popular multi-Higgs models. The simulation
of the SM background proceeds analogously, using instead the Standard Model
UFO library shipped with {\sc MadGraph}5\_{aMC@NLO}.

The simulation of the parton showering and hadronization is performed by means
of the {\sc Pythia}~6 program~\cite{Sjostrand:2006za}, and we include the
simulation of the response of a typical LHC detector as modeled by {\sc
Delphes}~3~\cite{deFavereau:2013fsa}, relying on the CMS-MA5tune
parameterization of the detector~\cite{Dumont:2014tja}. The resulting
detector-level objects are then reconstructed by applying the anti-$k_T$ jet
algorithm~\cite{Cacciari:2008gp}. More precisely, this is achieved by making use
of the {\sc MadAnalysis}~5 framework~\cite{Conte:2012fm,Conte:2014zja} to
simulate the detector effects and reconstruct the events (through an interface
to {\sc FastJet}~\cite{Cacciari:2011ma}), such a framework being also used to
implement the analyses described in the next subsections.

\subsection{Probing Higgs Cascades in the $4\ell2b$ final state}
\label{sec:4l2b}

\begin {table}
 \begin{center}
  \renewcommand{\arraystretch}{1.4}
  \setlength\tabcolsep{8pt}
  \begin{tabular}{c || c | c | c | c | c }
    Background & $ZZb\bar b$ & $t\bar t Z$ & $t\bar t W$ &
      $WWZb\bar b$ & $WWWb\bar b$\\
    \hline\hline
    $\sigma \times$ BR & 0.1~fb &1.2~fb & 2.3~fb & 1.2~fb & 2.1~fb\\
  \end{tabular}
  \caption{Leading-order cross section for the different SM processes
    contributing to the background of our $4\ell+2b$ analysis. They include the
    the relevant branching rations and the
    preselection cuts of Eq.~\eqref{eq:pres1} and Eq.~\eqref{eq:pres2}.}
  \label{table:bkgcs}
 \end{center}
\end{table}

In this section, we focus on the process of Eq.~\eqref{eq:process} when both
$Z$-bosons decay leptonically,
\be
 p p \to H_2 \to H_1 Z \to \h Z Z
     \to b \bar b\ \ell^+_1 \ell^-_1\ \ell^+_2 \ell^-_2 \ .
\label{eq:s1}\ee
The signal under consideration is thus made of one pair of $b$-jets and two
pairs of opposite-sign same-flavor leptons. In our analysis, we restrict
ourselves to lepton and jet candidates whose transverse momentum ($p_T^\ell$ and
$p_T^j$) and pseudoradipity ($\eta^\ell$ and $\eta^j$) satisfy
\be
  p_T^j > 20~{\rm GeV}, \qquad p_T^\ell > 10~{\rm GeV}, \qquad
  |\eta^j|<5\qquad\text{and}\qquad  |\eta^\ell| <2.5 \ .
\label{eq:pres1}\ee
Moreover, we omit from the analysis any pair of jet candidates that would not be
well separated from each other as well as any lepton that would be too close to
a jet.
In practice, we impose that the angular distance in the transverse plane between
two jets ($\Delta R_{jj}$) and the one between a jet and a lepton
($\Delta R_{\ell j}$) satisfy
\be
  \Delta R_{jj} > 0.4 \qquad\text{and}\qquad
  \Delta R_{\ell j} > 0.4 \ .
\label{eq:pres2}\ee
The dominant contributions to the SM background hence arise from $ZZb\bar b$,
$t\bar t V$ and $WWVb\bar b$ production, with $V$ being a $W$-boson or a
$Z$-boson. Including the
branching ratio (BR) corresponding to the $4\ell2b$ final state, the
leading-order cross sections as returned by {\sc MadGraph}5\_aMC@NLO are given
in Table~\ref{table:bkgcs}.

\begin{figure}
 \centering
  \includegraphics[width=0.48\columnwidth]{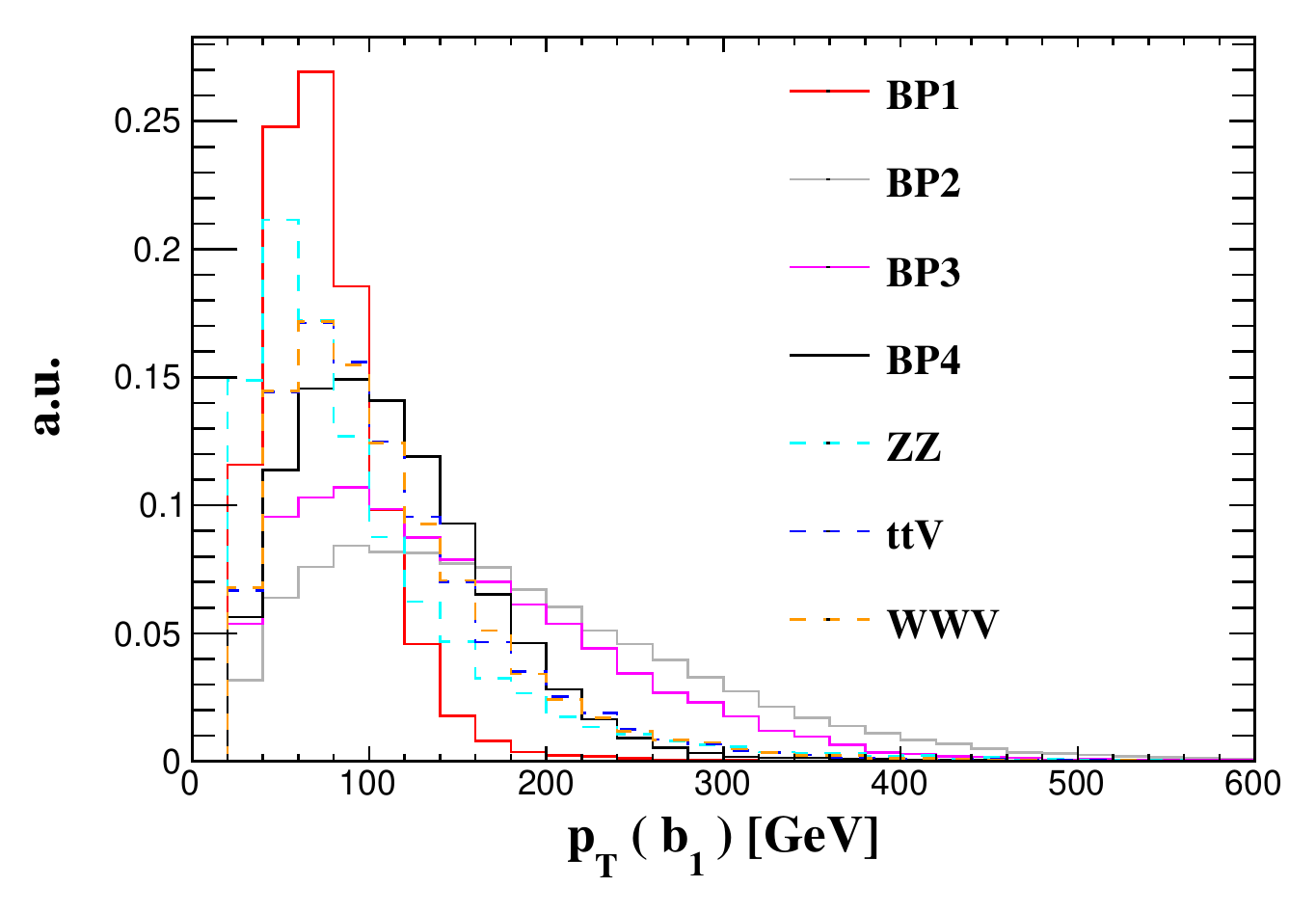}
  \includegraphics[width=0.48\columnwidth]{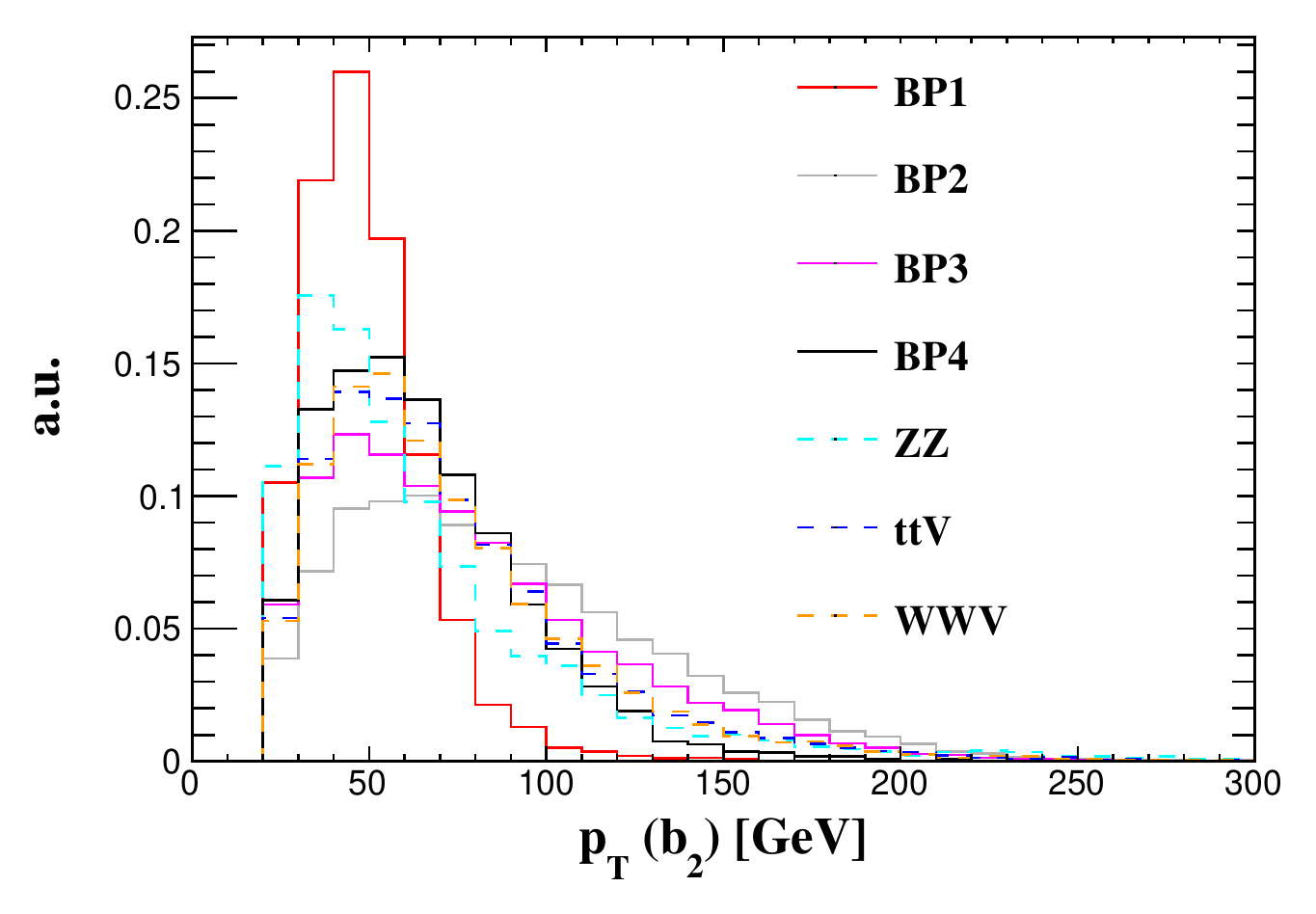}
  \caption {Normalised transverse-momentum distribution of the leading (left)
    and next-to-leading (right) $b$-tagged jets after having selected events
    featuring two pairs of charged leptons.}
  \label{fig:ptb}
 \includegraphics[width=0.49\columnwidth]{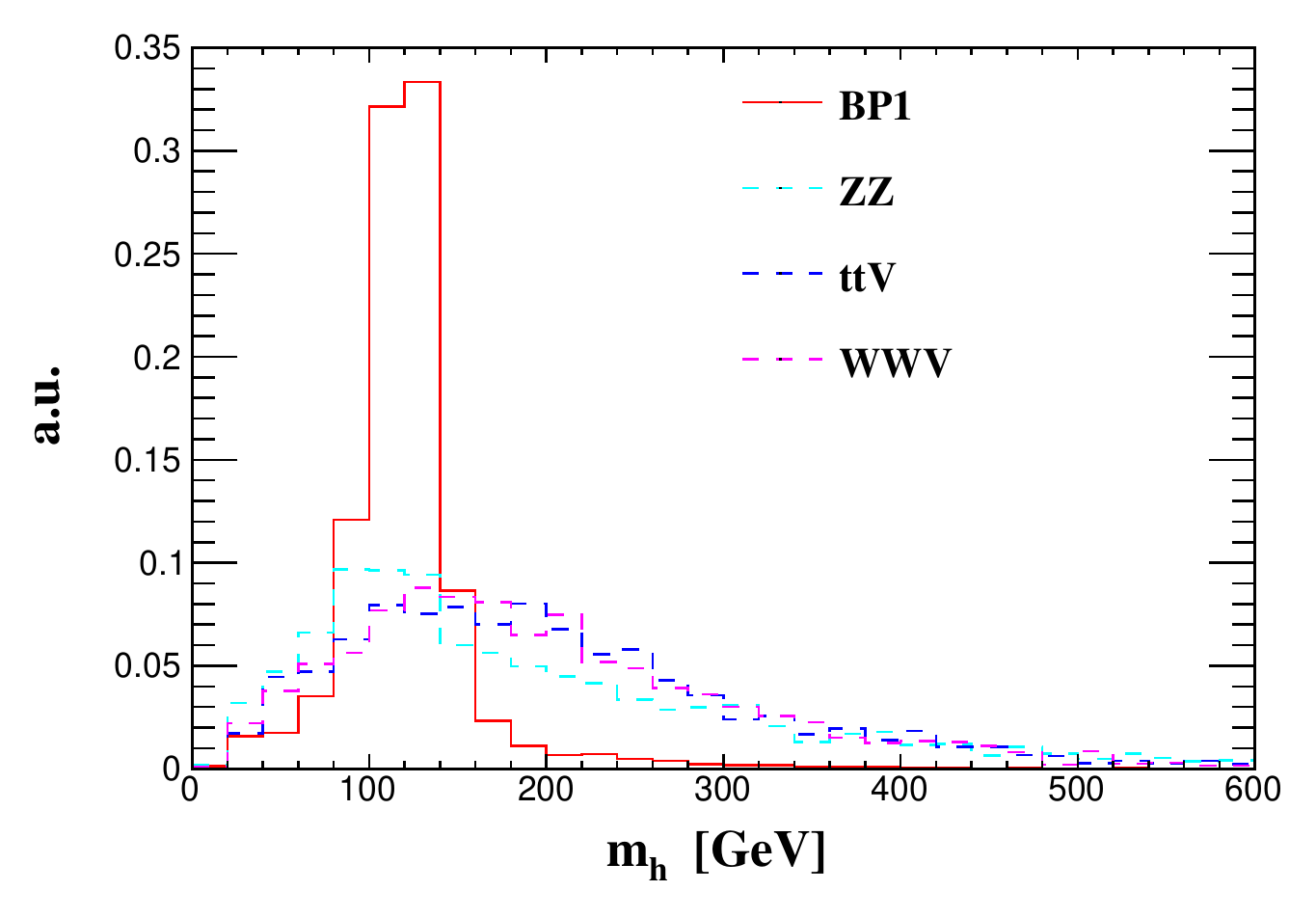}
  \includegraphics[width=0.49\columnwidth]{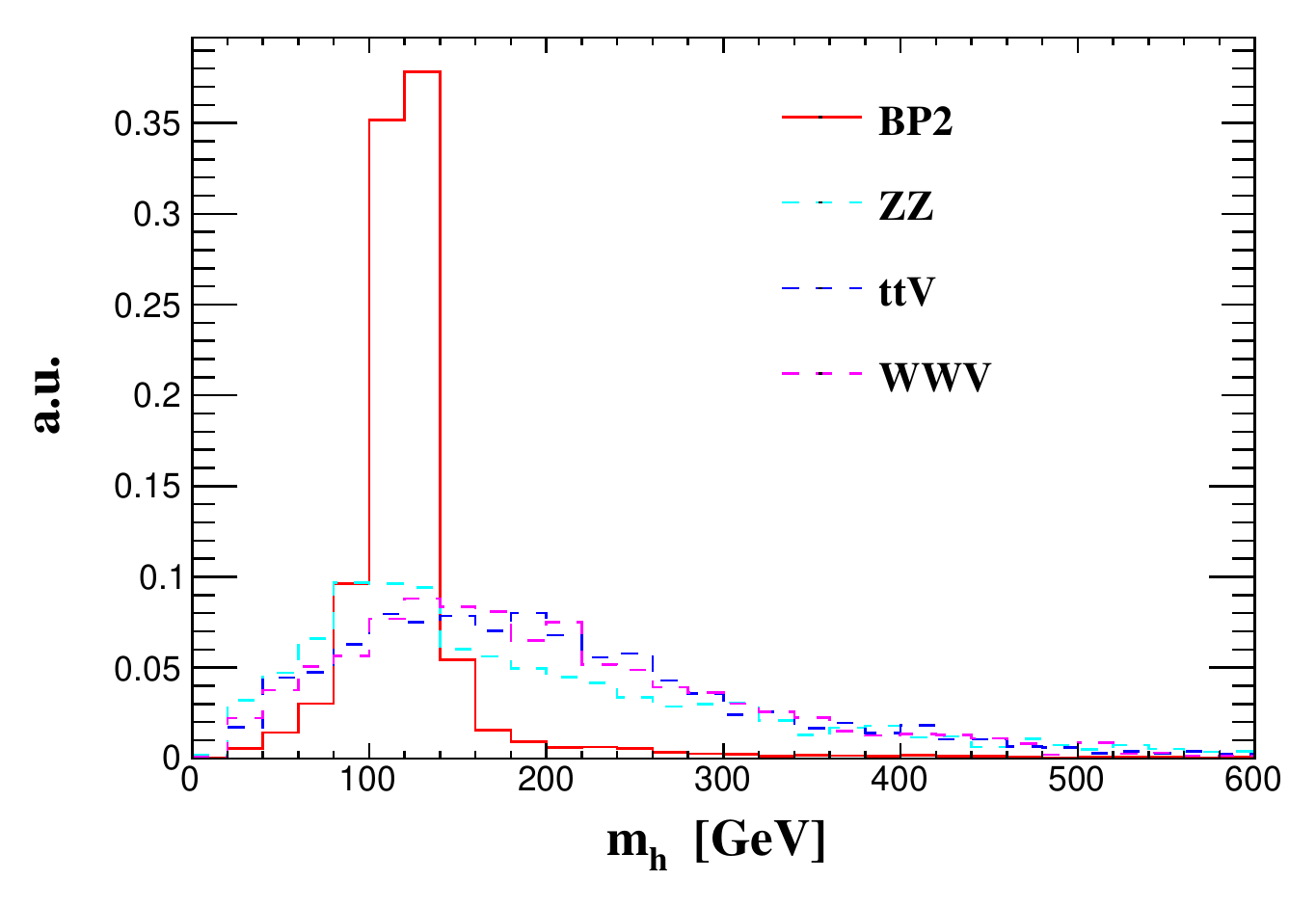}
 \caption{Normalized invariant-mass spectrum of the system comprised of the two
    leading $b$-jets,
    in the context of the $N(b)=2$ analysis. Results are shown for both the
    signal corresponding to the {\bf BP1} (left) and {\bf BP2} (right) scenarios,
    and for the dominant contributions to the background.}
 \label{fig:mbb-BP12}
  \includegraphics[width=0.49\columnwidth]
     {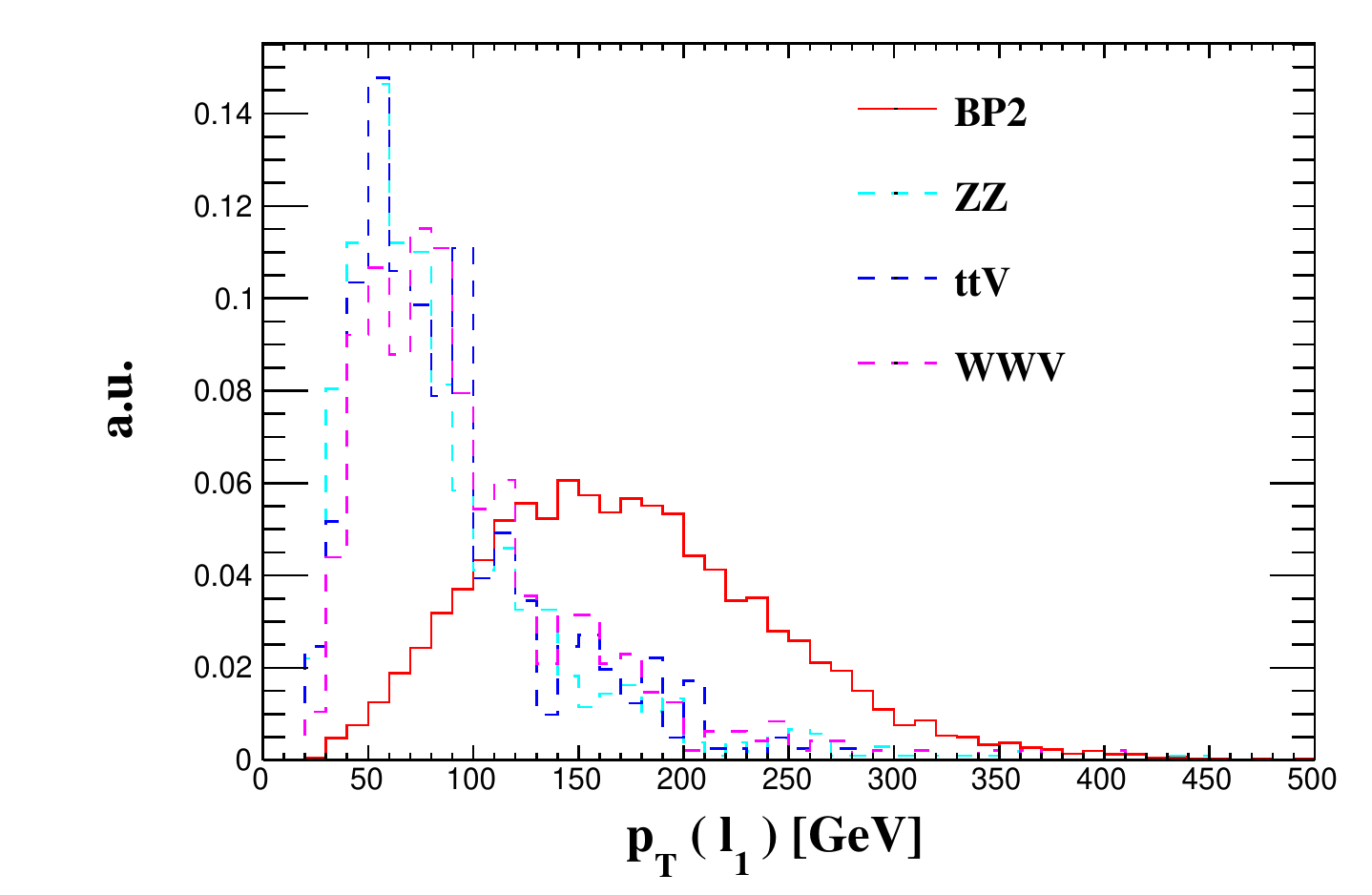}
  \includegraphics[width=0.49\columnwidth]
     {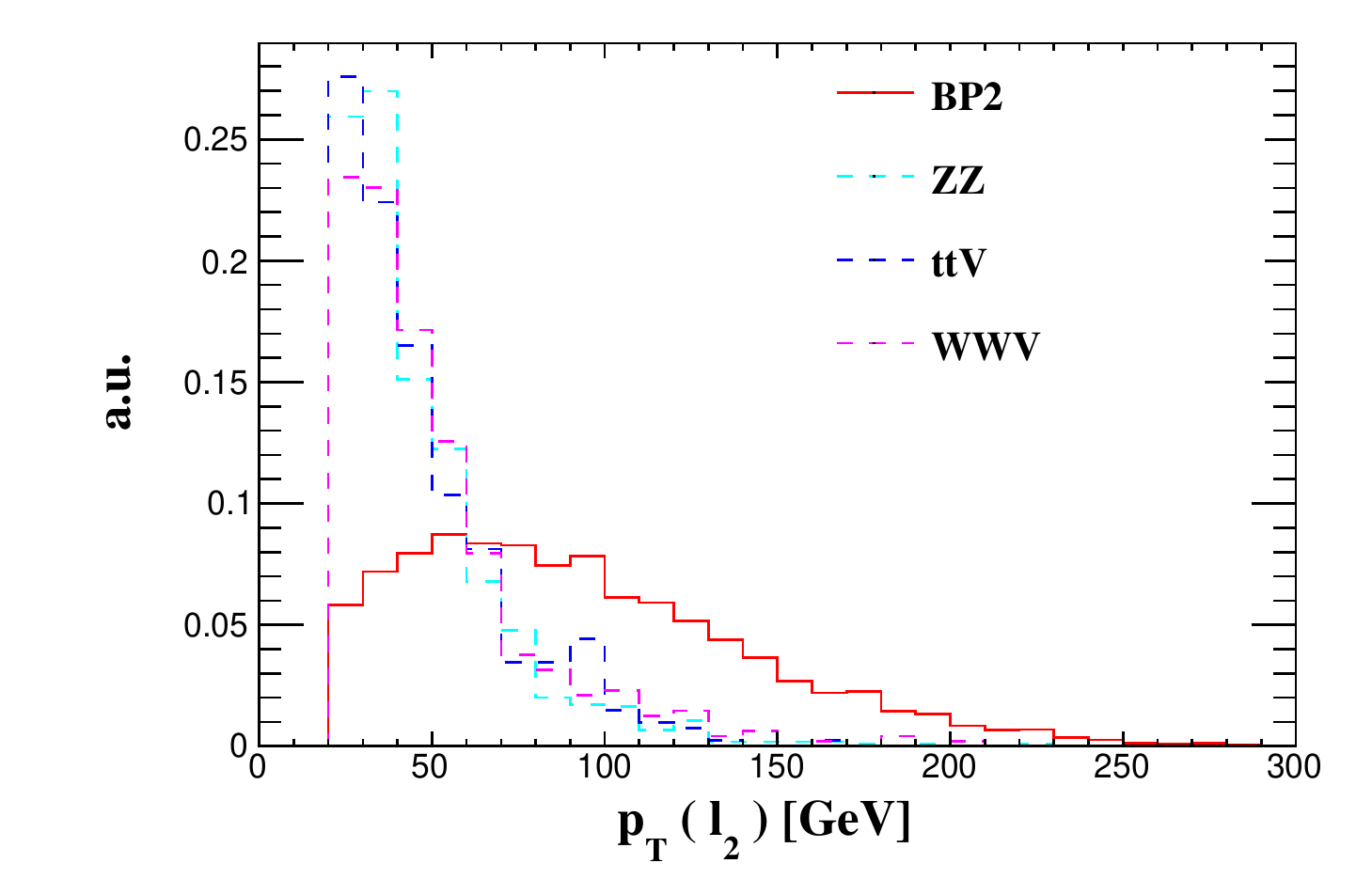}
 \caption{Normalised $p_T$ distributions of the leading (left) and
    next-to-leading (right) leptons, both for the signal corresponding to
    the {\bf BP2} scenario and for the dominant background contributions, in the
    case of the $N_b=2$ signal region.}
 \label{figure:bp2:1b4l:afterMbbcut}
\end{figure}

\begin{figure}
 \centering
  \includegraphics[width=0.48\columnwidth]{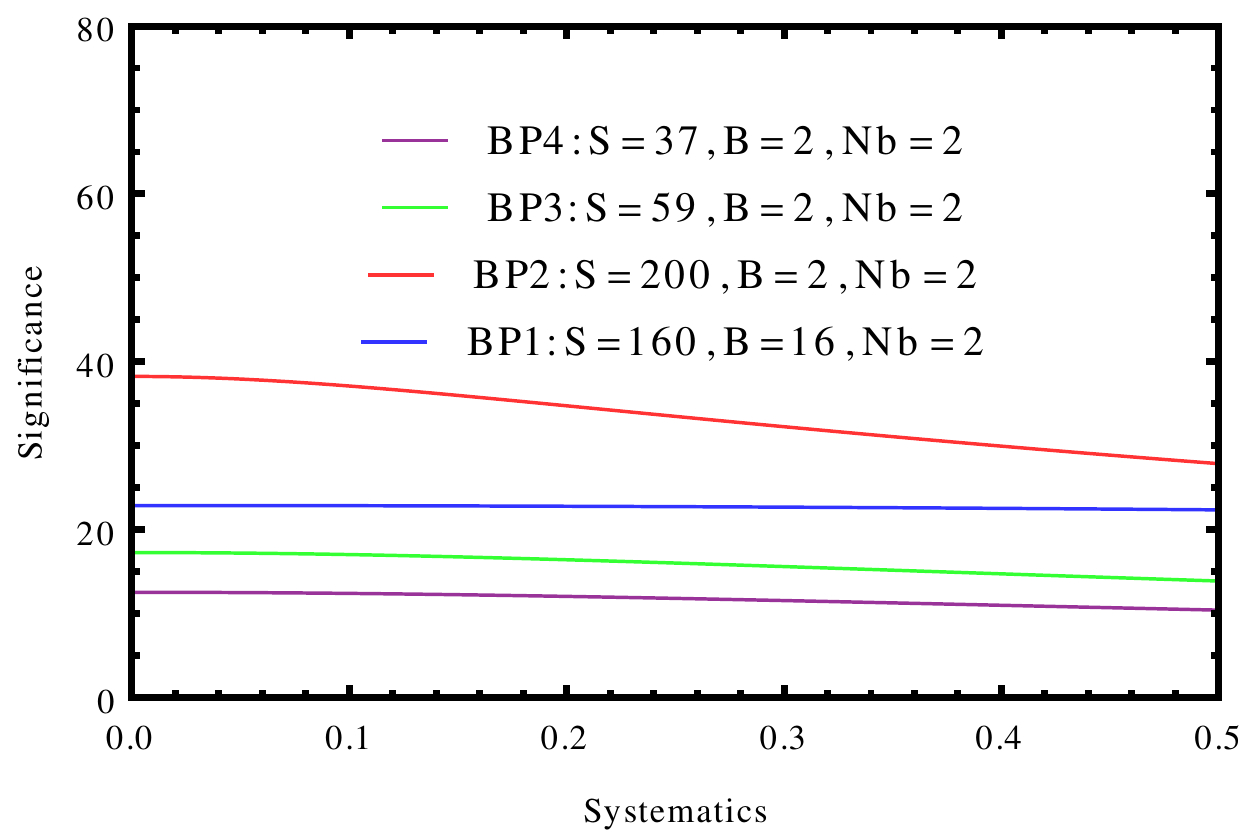}
  \includegraphics[width=0.48\columnwidth]{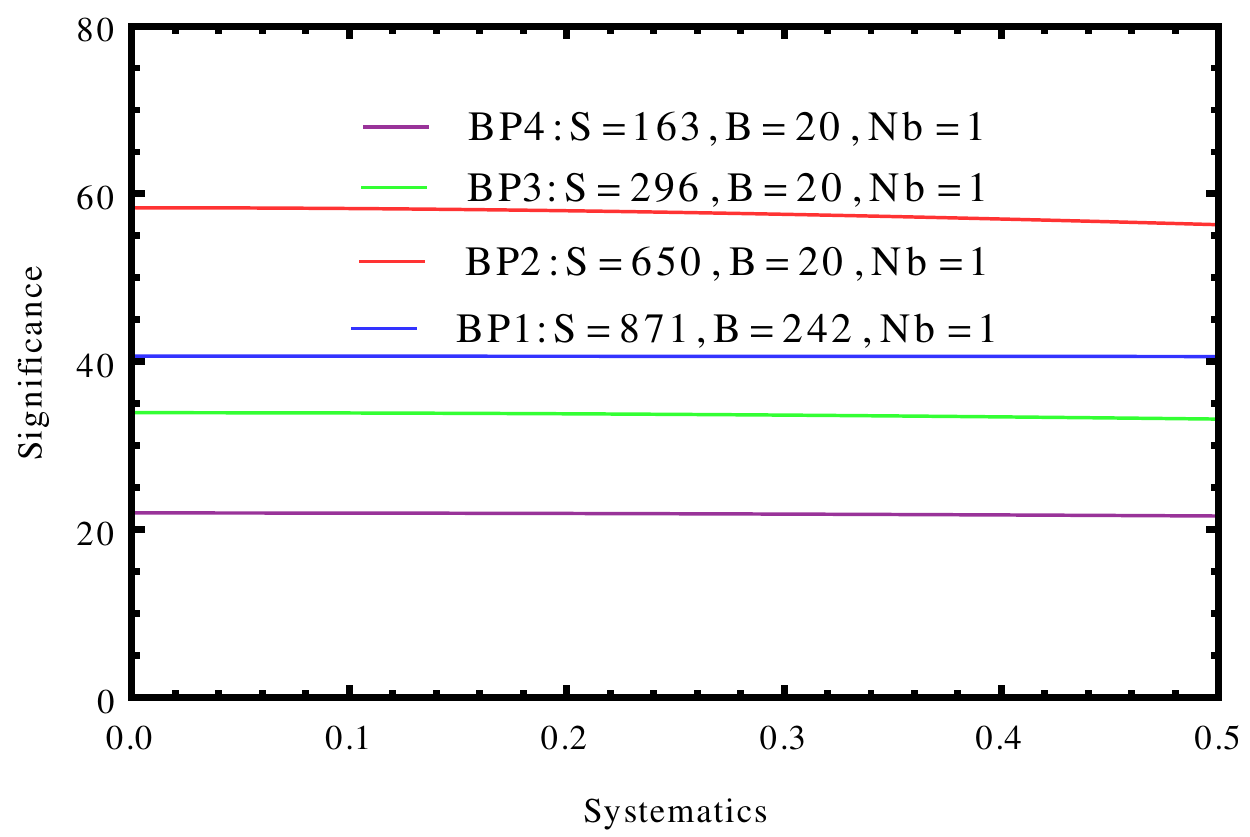}
  \caption{LHC significance, as defined by Eq.~\eqref{eq:sign},
    to the considered Higgs cascade decays for the four
    considered benchmark scenarios and assuming a luminosity of 1000~fb$^{-1}$.
    We show results for the $N(b)=2$ (left) and $N(b)=1$ (right) signal
    regions, and calculate the dependence of the significance on the level of
    systematic uncertainties taken as $\Delta_B = x B$ (with $x$ being shown on
    the $x$-axis).}
  \label{figure:signi_x}
  \includegraphics[width=0.48\columnwidth]{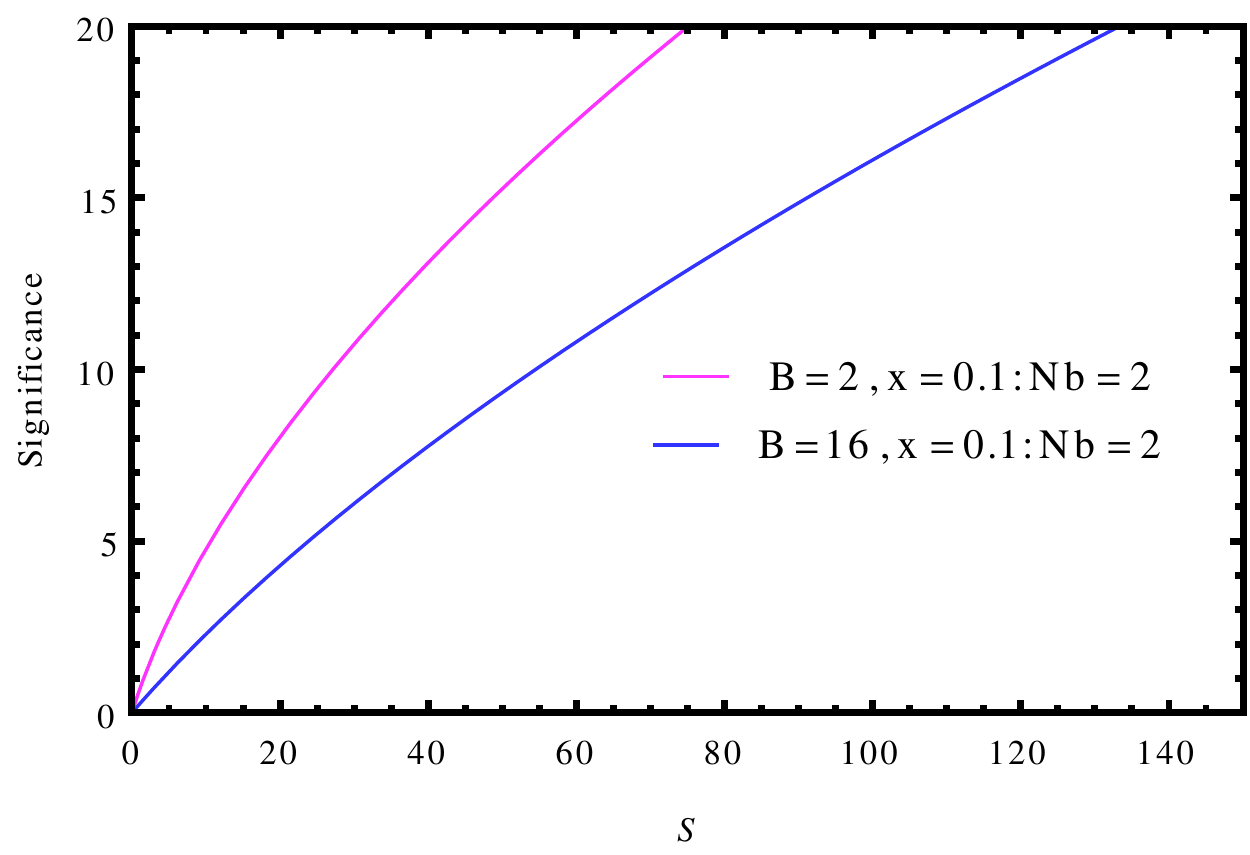}
  \includegraphics[width=0.48\columnwidth]{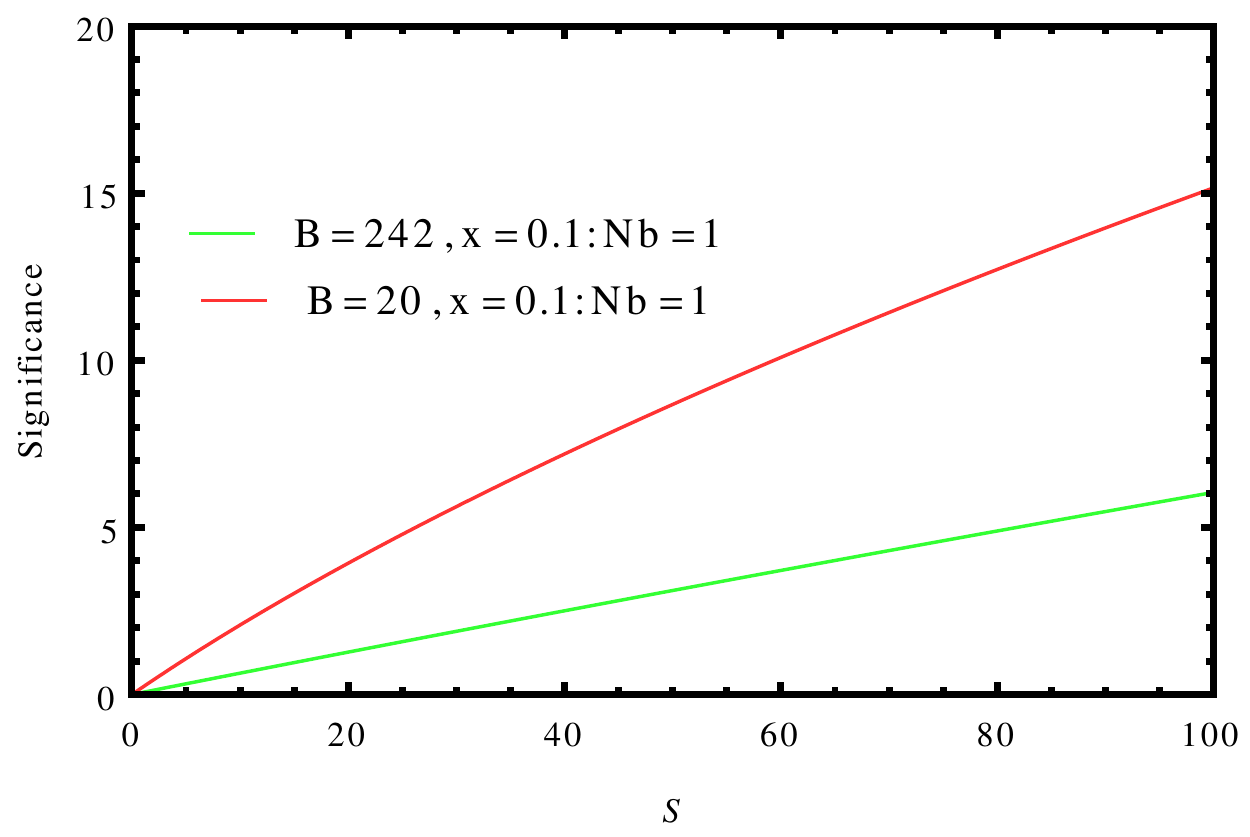}
  \caption{Variation of the significance with respect to the number of signal
    events $S$ for the $N(b)=2$ (left) and $N(b)=1$ (right) signal regions, both
    when the cut on the transverse momentum of the leading and next-to-leading
    leptons is applied (purple and red) and ignored (blue and green). We
    consider a level of systematic uncertainties of 10\%.}
  \label{figure:signi_S}.
\end{figure}

We implement a flavor-blind analysis in order to increase the signal statistics,
although we ignore tau leptons as those objects are more complicated to
reconstruct. We hence focus on leptons of the first two generations, so that
$\ell_1, \ell_2 = e, \mu$ in Eq.~\eqref{eq:s1}, and we require the presence of
two positively-charged and two negatively-charged leptons,
\be
  N(\ell^+) = N(\ell^-) = 2 \ .
\ee
The corresponding signal selection efficiency is about 40\%, many leptons
being missed as lying outside the acceptance of the detector or being too soft
for being correctly reconstructed. The corresponding background rejection
factor is slightly below 7, as many of the background components do not yield a
tetraleptonic signal.

Although the signal is expected to feature the presence of two $b$-jets,
$b$-tagging is not perfect. Harder $b$-jets are indeed more easily correctly
reconstructed than softer $b$-jets. The transverse momentum distributions of
the two leading $b$-tagged jets is illustrated in Figure~\ref{fig:ptb}, in which
we can observe that the bulk of the events feature softer $b$-jets.
Moreover, for scenarios where the mass splitting
between the Higgs bosons is large, the produced SM Higgs boson is often boosted.
The two $b$-jets are therefore not resolved, and a single $b$-jet is instead
reconstructed.
It consequently turns out that only 10--20\% of the surviving signal events
contain two tagged $b$-jets. For the
{\bf BP2} and {\bf BP4} scenarios, the mass splittings between the different
Higgs states is large and $b$-jets are more efficiently tagged, the signal
selection efficiency being larger. In contrast, the selection efficiency is
found to be smaller for the two other scenarios, as the $H_1\to \h Z$ decay
proceeds almost at threshold.

On the other hand,
40--50\% of the signal events are tagged as single-$b$-jet events, and a
significant fraction of them do not feature any tagged $b$-jets at all. In order
to recover the large number of signal events featuring a single $b$-jet, we
consider two independent signal regions in which we respectively require 2 and 1
$b$-tagged jet,
\be
  N(b)=2 \qquad\text{or}\qquad N(b)= 1 \ .
\ee
This cut allows for reducing the background by a factor of about 10 and 2.5 in
the two and one jet cases respectively.

\begin{table}
 \begin{center}
  \renewcommand{\arraystretch}{1.4}
  \setlength\tabcolsep{8pt}
  \begin{tabular}{c| c ||c|c|c|c||c}
    \multicolumn{2}{c||}{Selection step} & {\bf BP1} & {\bf BP2} & {\bf BP3} &
       {\bf BP4} & Background\\
    \hline\hline
    0 & Initial                   & 5000 & 5000 & 5000 & 5000 & 13636\\ \hline
    1 & $N(\ell^+)=N(\ell^-)=2$   & 1993 & 2723 & 1979 & 2373 &  1992\\ \hline
    2 & $N(b)=2$                  &  206 &  490 &  260 &  340 &   231\\ \hline
    3 & $\slashed{E}_T<50$~GeV    &  203 &  415 &  220 &  321 &    66\\ \hline
    4 & 90~GeV $<M_{bb}<$ 150~GeV &  160 &  344 &  174 &  257 &    16\\ \hline
    \multirow{2}{*}{5} & $p_T(\ell_1)>75$ GeV & \multirow{2}{*}{$\sim 0$} &
       \multirow{2}{*}{200} & \multirow{2}{*}{59} & \multirow{2}{*}{37} &
       \multirow{2}{*}{2}\\
    & $p_T(\ell_2) > 50$~GeV&&&&&\\
  \end{tabular}\\[.2cm]
  \begin{tabular}{c| c ||c|c|c|c||c}
    \multicolumn{2}{c||}{Selection step} & {\bf BP1} & {\bf BP2} & {\bf BP3} &
       {\bf BP4} & Background\\
    \hline\hline
    0 & \hspace{1.45cm}Initial\hspace{1.45cm}
                                  & 5000 & 5000 & 5000 & 5000 & 13636\\ \hline
    1 & $N(\ell^+)=N(\ell^-)=2$   & 1993 & 2723 & 1979 & 2373 &  1992\\ \hline
    2 & $N(b)=1$                  &  884 & 1310 &  910 & 1115 &   818\\ \hline
    3 & $\slashed{E}_T<50$~GeV    &  871 & 1122 &  782 & 1060 &   242\\ \hline
    \multirow{2}{*}{4} & $p_T(\ell_1)>75$ GeV & \multirow{2}{*}{$\sim 0$} &
       \multirow{2}{*}{650} & \multirow{2}{*}{296} & \multirow{2}{*}{163} &
       \multirow{2}{*}{20}\\
    & $p_T(\ell_2) > 50$~GeV&&&&&\\
  \end{tabular}
  \caption{Number of events surviving each selection step for the four
    considered benchmark scenarios, as well as for the SM background. The
    results are normalized to an integrated luminosity of 1000~fb$^{-1}$ and
    include a conservative $K$-factor of 2 for the background. Results are
    presented for the $N(b)=2$ signal region (upper table) and $N(b)=1$ signal
    region (lower table).}
  \label{tab:cutflows}
 \end{center}
\end{table}

As shown in Table~\ref{tab:cutflows} for an integrated luminosity of
1000~fb$^{-1}$, about 200--500 and 900--1300 signal events
are expected to respectively populate the $N(b)=2$ and $N(b)=1$ signal regions,
to be compared with 230 and 800 background events (including a conservative
$K$-factor of 2). From this stage,
background rejection can be improved by restricting the missing transverse
energy $\slashed{E}_T$ in the event,
\be
  \slashed{E}_T < 50~{\rm GeV}.
\ee
This selection leaves the signal barely unaffected as it is expected to be
fully visible, and reduces the background by an extra factor of 3. The surviving
background events are mostly originating from $t\bar t Z$ and $WWZb \bar b$
production. In the $N(b)=2$ signal region, an extra selection can be imposed as
the invariant mass of the dijet system $M_{bb}$ has to be compatible with the
mass of the Higgs boson,
\be
  90~{\rm GeV} < M_{bb} < 150~{\rm  GeV} \ .
\ee
The importance of this last selection is demonstrated in
Figure~\ref{fig:mbb-BP12}
for the {\bf BP1} and {\bf BP2} scenarios, where the distributions in the
invariant mass of the system made of the two leading $b$-tagged jets is shown for
two representative benchmark scenarios, and the main contribution to the
background. The availability of reconstructing the Standard Model Higgs boson is
hence crucial when searching for heavier Higgs bosons, and achieveable even for
compressed spectra.

Finally, we make use of the different properties of the leading lepton $\ell_1$
and next-to-leading lepton $\ell_2$ for the signal and the background (as
illustrated in Figure~\ref{figure:bp2:1b4l:afterMbbcut} for the {\bf BP2}
scenario) to further improve the senstivity, enforcing
\be\label{eq:ptl}
  p_T(\ell_1) > 75~{\rm GeV} \qquad\text{and}\qquad p_T(\ell_2) > 50~{\rm GeV.}
\ee
These two last cuts yield a basically background-free environment. The
corresponding signal selection efficiencies are usually large, except for
scenarios featuring a small mass splitting such as in the {\bf BP1}
configuration. We will therefore ignore this cut for
what concerns the {\bf BP1} configuration.

In addition to the conservative $K$-factor of 2 that has been included in the
background numbers to model higher-order effects, we assess the potential
effects of the systematic
uncertainties of $x\%$ by computing the signal significance as~\cite{%
Cowan:2010js}
\be
 Z = \sqrt{2}\Bigg(
  (S+B)~\ln\left[\frac{(S+B)~(B+\Delta_B^2)}{B^2+(S+B)~\Delta_B^2}\right]
      - \frac{B^2}{\Delta_B^2}\ln\left[1+\frac{\Delta_B^2~S}{B~(B+\Delta_B^2)}
         \right]\Bigg)^\frac12
  \qquad\text{with}\qquad\Delta_B = x B \ .
\label{eq:sign}\ee
In Figure~\ref{figure:signi_x}, we present the LHC sensitivity, as defined by
Eq.~\eqref{eq:sign}, to the signal for the different considered benchmark
scenarios and for both the $N(b)=2$ and $N(b)=1$ signal regions. The
normalization moreover corresponds to an integrated luminosity of
1000~fb$^{-1}$.
All the selection cuts introduced above have been applied, with the exception of
the one on the leading and next-to-ledaing leptons in the case of the {\bf BP1}
scenario, as indicated above. The results are shown for various levels of
systematic uncertainties ranging from 0\% to 50\%. They are found stable with
respect to the systematics thanks to a very large signal and the almost
background-free environment originating from our selection.

The results presented so far correspond to a signal cross section that has been
fixed to 5~fb. In Figure~\ref{figure:signi_S}, we relax this hypothesis and
show the dependence of the significance on the number of signal events $S$
when 10\% of systematic uncertainties is assumed. We again consider
both the $N(b)=2$ (left) and $N(b)=1$ (right) signal regions, and study the
dependence on the last cut on the transverse momentum of the
two leading leptons. As expected, the effect of this selection increases the
significance for a given number of signal events. Conversely, while about 25 and
85 signal events are required for a $5\sigma$ discovery without imposing any
requirement on the leptons, for the $N(b)=2$ and $N(b)=1$ signal regions
respectively, these numbers are reduced to 10 and 25 after constraining the
transverse momentum of the leptons as in Eq.~\eqref{eq:ptl}.

\begin{figure}
 \centering
  \includegraphics[width=0.49\columnwidth]
    {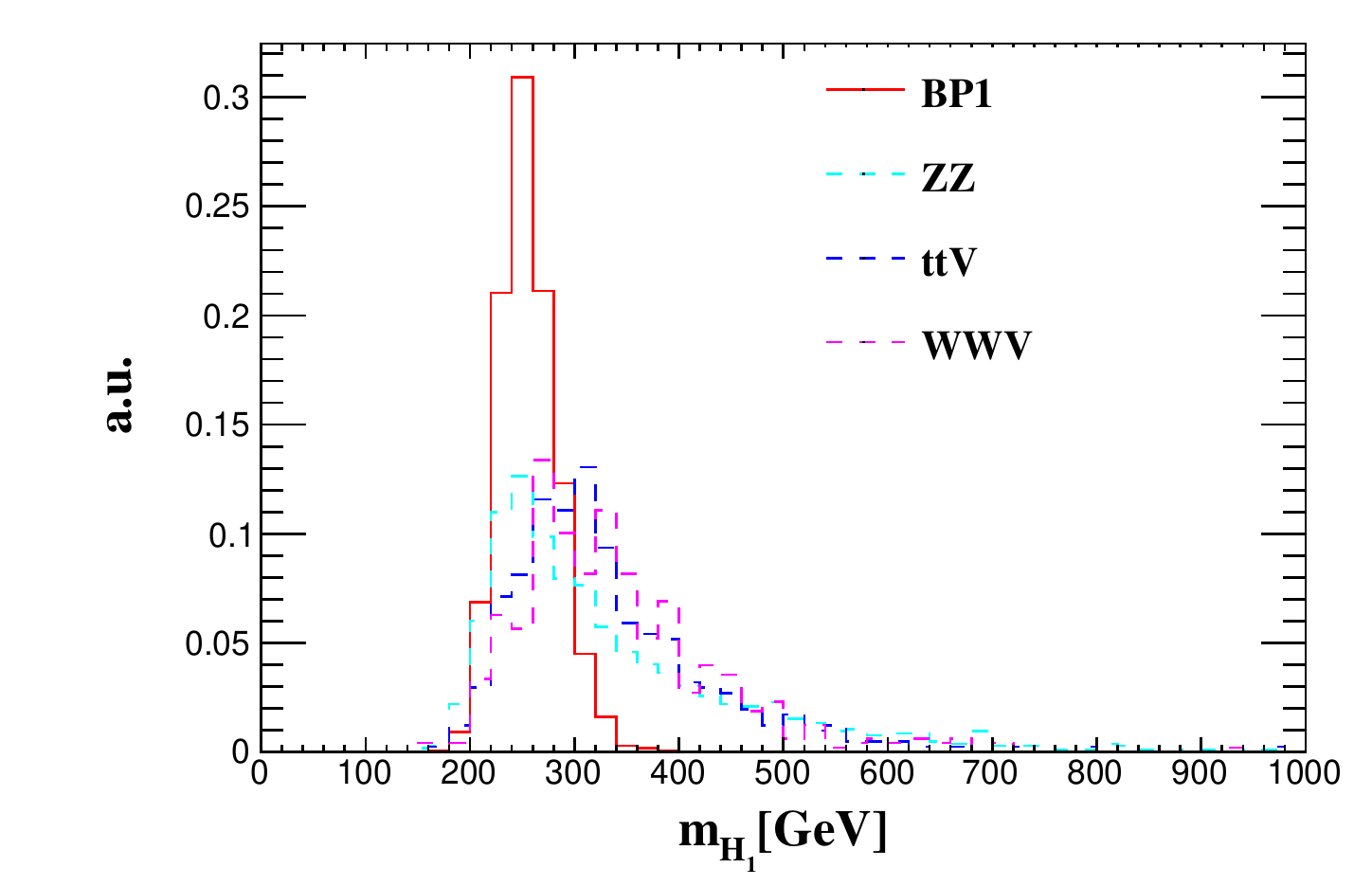}
  \includegraphics[width=0.49\columnwidth]
    {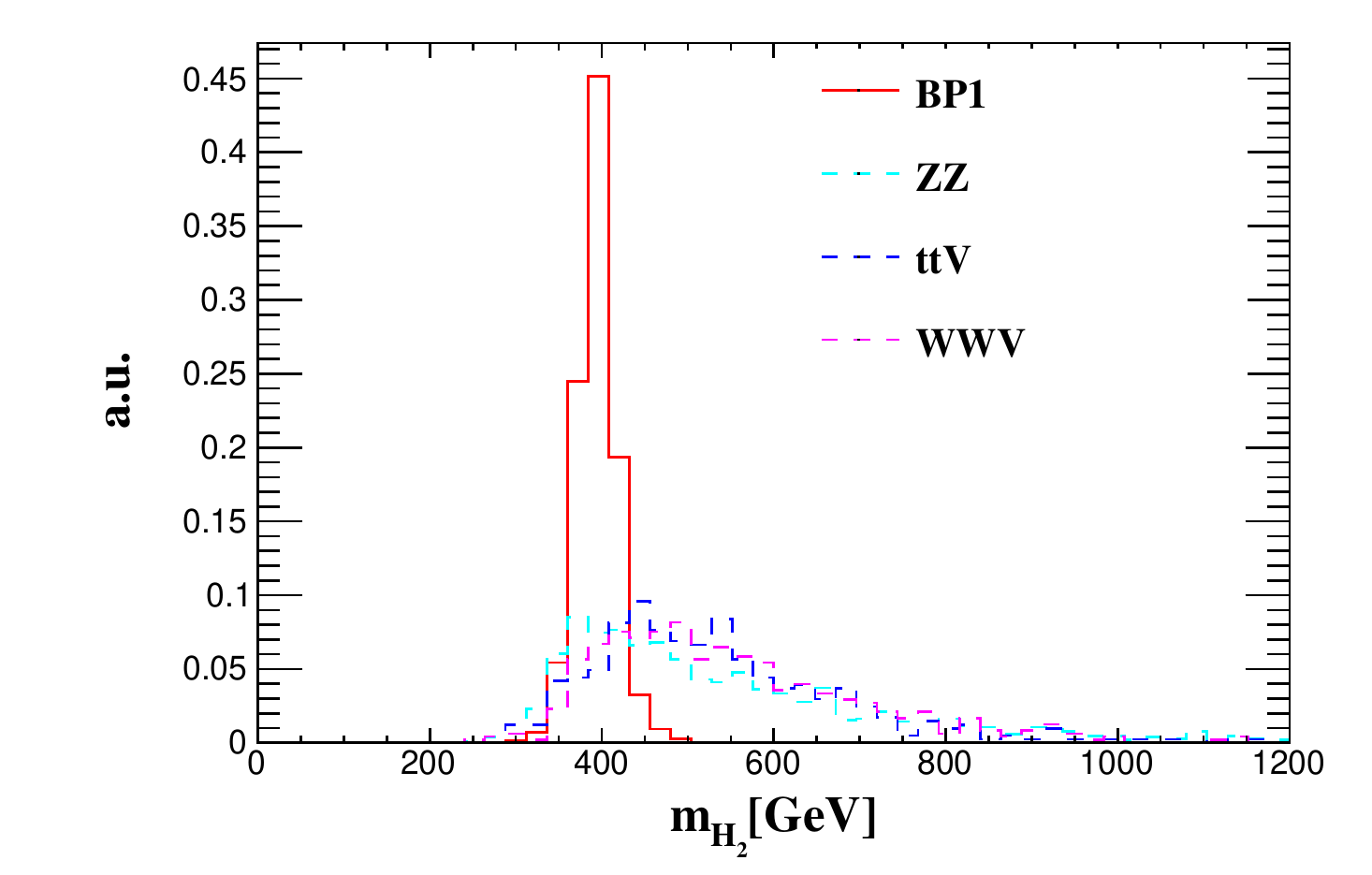}
  \caption{Normalized invariant-mass spectrum for the intermediate $H_1$ (left)
    and
    $H_2$ states in the context of the $N(b)=2$ analysis (the last cut being
    omitted) and for the {\bf BP1} scenario. The results are derived from the
    reconstruction of the $2b\ell^+\ell^-$ and $2b4\ell$ systems. Whilst the
    spread in the $H_1$ invariant mass spectrum stems from the different
    possible combinations of the leptons, the distribution is found similar to
    the one obtained for any other lepton combination.}
  \label{fig:bp1:recon2b}
\end{figure}

Translating these numbers in term of cross section, compressed scenarios like
our {\bf BP1} configuration could yield an observable signal in the $N(b)=2$
and $N(b)=1$ signal regions as long as the production rate is at least
0.78~fb and 0.49~fb, respectively. The reach of the $N(b)=1$ signal region is
found to be larger, by virtue of the efficiency to correctly identify one
$b$-jet that is larger than the one to identify two $b$-jets. The two regions
are however complementary, as even if the $N(b)=1$ region is better for what
concerns the reach, the $N(b)=2$ analysis offers a way to uniquely reconstruct
the intermediate heavy Higgs states as illustrated in
Figure~\ref{fig:bp1:recon2b}. For scenarios exhibiting a mass spectrum featuring
larger splittings like in the {\bf BP2} configuration, the final-state objects
are harder, which implies a better reconstruction efficiency. Accordingly, one
obtains better expected limits on the production rate, the observable cross
section being 0.25~fb and 0.19~fb in the $N(b)=2$ and $N(b)=1$ analysis,
respectively. For scenarios featuring a smaller mass splitting like in the
{\bf BP3} case (where one of the decays has to occur close to threshold) or in
the {\bf BP4} case (where both decays have less available phase space than in
the {\bf BP2} case), the
final-state objects are softer, which results in degraded expected limits on the
signal cross section, but still in the 1~fb regime.

\subsection{Probing Higgs Cascades in the $2\ell4b$ final state}
\begin {table}
 \begin{center}
  \renewcommand{\arraystretch}{1.4}
  \setlength\tabcolsep{8pt}
  \begin{tabular}{c || c | c | c }
    Background & $\ell^+\ell^-$ + jets & $W^+ W^-$ + jets & $t\bar t \h$\\
    \hline\hline
    $\sigma\times$ BR & 3.2~pb & 109.1~fb & 14~fb
  \end{tabular}
  \caption{Leading-order cross section for the different SM processes
    contributing to the background of our $2\ell+4b$ analysis. They include the
    the relevant branching ratios and the
    preselection cuts of Eq.~\eqref{eq:pres1} and Eq.~\eqref{eq:pres2}.}
  \label{table:bkgcs2l4b}
 \end{center}
\end{table}

The Higgs cascade signal that we consider in this work could also give rise to a
final-state signature comprised of four $b$-jets and one pair of opposite-sign
leptons of the same flavor,
\be
 p p \to H_2 \to H_1 Z \to \h Z Z \to b \bar b\ b \bar b\ \ell^+ \ell^- \ .
\ee
The combinatorics induced by the final-state recontruction and the more abundant
SM background renders the task of discriminating the signal from the background
complicated. We however verify, in this section, the existence of any potential
corresponding handle. The dominant contributions to the SM background arise
from the associated
production of a Drell-Yan pair of leptons with jets, $W$-boson pair production
with jets and $t\bar t \h$ production. The leading-order cross sections as
returned by {\sc MadGraph}5\_aMC@NLO are shown in Table~\ref{table:bkgcs2l4b}.

\begin{table}
 \begin{center}
  \renewcommand{\arraystretch}{1.4}
  \setlength\tabcolsep{8pt}
  \begin{tabular}{c| c ||c|c|c|c||c}
    \multicolumn{2}{c||}{Selection step} & {\bf BP1} & {\bf BP2} & {\bf BP3} &
       {\bf BP4} & Background\\
    \hline\hline
    0 & Initial                   & 5000 & 5000 & 5000 & 5000 & $6.657 \times 10^6$\\ \hline
    1 & $N(\ell^+)=N(\ell^-)=1$   & 2815 & 3006 & 2747 & 2971 & $3.695 \times 10^6$\\ \hline
    2 & $N(j)\ge 4$               & 2811 & 3004 & 2735 & 2970 & $3.644 \times 10^6$\\ \hline
    3 & $N(b)= 3$                 &  228 &  506 &  302 &  394 &   25062\\ \hline
    4 & 80~GeV$<M_{\ell\ell}<$100~GeV
                                  &  201 &  434 &  258 &  343 &   13072\\ \hline
    \hline
    5a & 300~GeV$<M_{H_2}<$500~GeV  & 121 & --  & -- & -- & 1954 \\ \hline
    5b & 900~GeV$<M_{H_2}<$1400~GeV, $p_T^\ell > 70$~GeV & --  & 192 & -- & -- &  455 \\ \hline
    5c & 900~GeV$<M_{H_2}<$1400~GeV, $p_T^\ell > 60$~GeV & --  & --  & 94 & -- &  649 \\ \hline
    5d & 500~GeV$<M_{H_2}<$700~GeV , $p_T^\ell > 50$~GeV & --  & --  & -- & 91 &  552 \\ \hline
  \end{tabular}
  \caption{Number of events surviving each selection step for the four
    considered benchmark scenarios, as well as for the SM background. The
    results are normalized to an integrated luminosity of 1000~fb$^{-1}$ and
    include a conservative $K$-factor of 2 for the background.}
 \label{tab:cflow2}
 \end{center}
\end{table}

In our analysis, jet and lepton candidates are selected as in
Eq.~\eqref{eq:pres1} and Eq.~\eqref{eq:pres2}. We preselect events
containing one positively-charged and one negatively-charged lepton regardless
of the lepton flavor,
\be
  N(\ell^+) = N(\ell^-) = 1 \ ,
\ee
and we require in addition the presence of at least four jets out of which three
should be $b$-tagged,
\be
  N(j) \geq 4 \qquad\text{with}\qquad N(b)=3 \ .
\ee
Whilst four $b$-tagged jets are expected, the loss in signal efficiency induced
by the requirement of a fourth $b$-tag would make the signal unobservable
(see the discussion in Section~\ref{sec:4l2b}). The combined signal
efficiency for these preselection cuts is of about 5--6\%, for a background
rejection factor of about 250. We then impose the lepton pair to be
compatible with the decay of a $Z$-boson, constraining its invariant mass
$M_{\ell\ell}$ to satisfy
\be
  80~{\rm GeV} \leq M_{\ell\ell} \leq 100~{\rm GeV}.
\ee
This allows for the reduction of the diboson and Higgs backgrounds without
impacting the signal too much. At this stage, the number of background events is
of about 13000, while the number of signal events is expected to be in the
200--450 window for the different scenarios, as illustrated in
Table~\ref{tab:cflow2}. The signal is thus
not visible over the background.

\begin{figure}
  \centering
  \includegraphics[width=0.65\columnwidth]
    {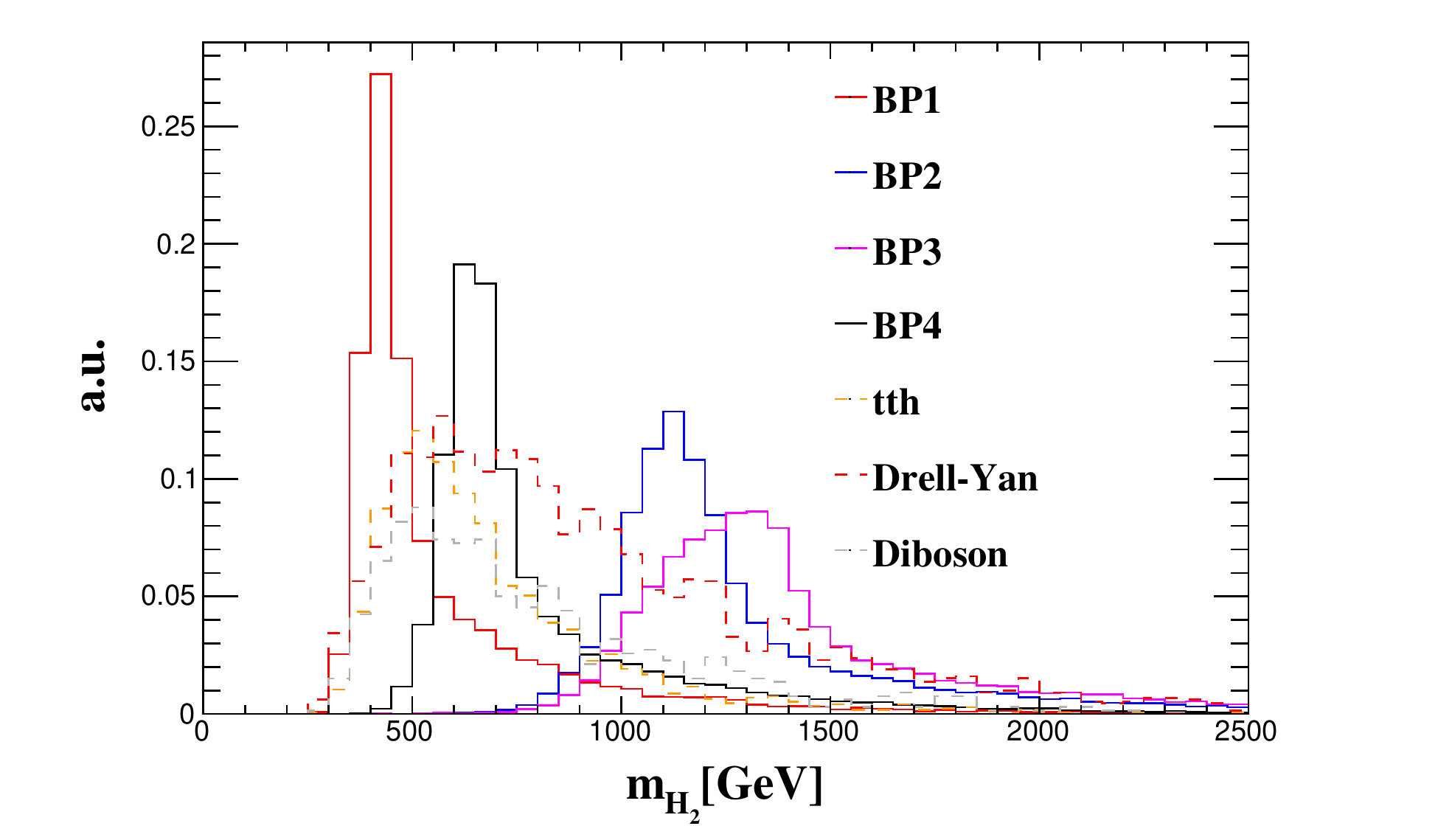}
  \caption {Normalised invariant-mass distribution of the heavy $H_2$ boson for
    the main contributions to the SM background and the four signal scenarios,
    once all other selection cuts have been applied.}
 \label{figure:4b2l_MpT}
\end{figure}

The fact that all final-state objects are not fully identified and the combinatorics that
may result from the reconstruction of the intermediate $H_1$ and $Z$-bosons make
a kinematic fit complicated, in particular once detector effects are accounted
for. We therefore approximate the invariant-mass
spectrum of the $H_2$ boson by the invariant-mass distribution of the system
comprised of the four leading jet candidates and the selected pair of leptons,
$M_{H_2}$. As illustrated in Figure~\ref{figure:4b2l_MpT}, this variable
serves as a good discriminator of the signal from the background. The
distribution turns to be very broad for some scenarios, the distortion being
larger for new physics scenarios featuring larger mass splittings as this
configuration could induce extra radiation and thus more jets in the final
state. There is no perfect scenario-independent selection that would allow for
the observation of the signal from the overwhelming background. Such a potential
cut indeed strongly depends on the mass splittings between the different Higgs
states. We therefore propose four different cuts,
\be\bsp
  ({\rm 5a}) &\hspace*{1cm}  300~{\rm GeV} < M_{H_2} <  500~{\rm GeV}\ , \\
  ({\rm 5b}) &\hspace*{1cm}  900~{\rm GeV} < M_{H_2} < 1400~{\rm GeV},
      \quad p_T^\ell > 70~{\rm GeV}\ , \\
  ({\rm 5c}) &\hspace*{1cm}  900~{\rm GeV} < M_{H_2} < 1400~{\rm GeV},
      \quad p_T^\ell > 60~{\rm GeV}\ , \\
  ({\rm 5d}) &\hspace*{1cm}  500~{\rm GeV} < M_{H_2} <  700~{\rm GeV},
      \quad p_T^\ell > 50~{\rm GeV}\ ,
\esp\ee
where the extra selection on the leptons allow for a better signal
discrimination in the case of a not too light spectrum (as this yields harder
leptons).
The first of these selection target setups similar to the configuration of the
{\bf BP1} scenario where the spectrum is compressed and light, while the second
selection aims for scenarios featuring heavier Higgs boson with enough mass
splittings to guarantee the presence of very hard leptons in the final state.
The third choice is also appropriate for heavier spectra, but it potentially
allows for intermediate decays being close to threshold. Finally, the last
selection targets spectra where the Higgs bosons are not too heavy but where the
decays can occur far from threshold.

For an integrated luminosity of 1000~fb$^{-1}$, these cuts lead to an LHC
sensitivity to the {\bf BP1}, {\bf BP2}, {\bf BP3} and {\bf BP4} scenarios of
$2.7\sigma$, $8.5\sigma$, $3.6\sigma$ and $3.8\sigma$ respectively, when 10\%
of systematic uncertainties is also factored in. These results are however found not
to depend on the systematics. Although potentially promising, the
$4b2\ell$ signature does not provide as clear a handle on the signal as the $2b4\ell$ channel and will therefore be not considered in what follows.

\section{Model Implications}
\label{sec:model}

We now turn to the understanding of the implications of the analyses
that have been designed in Section~\ref{sec:indep} in a simplified model
context. We investigate below how the simplified spectra introduced
in the previous section can be realized in a concrete model with an enlarged
scalar spectrum, and investigate the reach of our analysis.
As an operating example, we choose the Type-II 2HDM. For details about
the model and the couplings, we refer to Ref.~\cite{Branco:2011iw} and to
Section~\ref{sec:2hdm_basis} where we sketch the essential details. Our
phenomenological results are given in Section~\ref{sec:2hdm_results}.

\subsection{The Two-Higgs-Doublet Model - Spectrum and Couplings}
\label{sec:2hdm_basis}
The 2HDM has been extensively studied during the last decades, both as a standalonge model and also often as the
scalar sector of a larger model like the Minimal Supersymmetric Standard Model (MSSM). Unlike the SM, the 2HDM contains two weak
doublets of Higgs fields $\phi_1$ and $\phi_2$ of opposite hypercharge
$Y=\pm 1/2$. At the minimum of the potential, the neutral components of both
doublets develop vacuum expectation values (vev),
\be
  \langle \phi_1^0 \rangle = \frac{1}{\sqrt{2}} v_1
  \qquad\text{and}\qquad
  \langle \phi_2^0 \rangle = \frac{1}{\sqrt{2}} v_2 \ ,
\ee
where the vev of the SM Higgs fields $v$ is obtained
through $v_1^2+v_2^2 \equiv v^2 = (\sqrt{2} G_F)^{-1}$ with $G_F$ being the
Fermi constant. The two vevs $v_1$ and $v_2$ are thus not
arbitrary as their quadratic sum is connected to the mass scale of the
electroweak bosons. We have thus here a single free parameter that is often
chosen as the ratio $v_2/v_1=\tan\beta$.

The breaking of the electroweak symmetry induces a mixing of the
scalar degrees of freedom that reads, at tree-level,
\be
  \bpm \H\\ \h \epm =
    \bpm\phantom{-}\cos\alpha &\sin\alpha\\ -\sin\alpha&\cos\alpha\epm
    \bpm\Re\{\phi_1^0\}\\ \Re\{\phi_2^0\}\epm
  \ , \qquad
    \A =  - \sin\beta\ \Im\{\phi_1^0\} +\cos\beta\ \Im\{\phi_2^0\}
  \ , \qquad
  H^\pm = - \sin\beta\ \phi_1^\pm + \cos\beta\ \phi_2^\pm \ ,
\label{eq:mass}\ee
where $\h$ and $\H$ are $CP$-even mass-eigenstates, $\A$ is a $CP$-odd
mass-eigenstate and $H^\pm$ are the physical charged Higgs bosons. In the
notation of Section~\ref{sec:indep}, the $H_2$ boson can in principle equally
be mapped to the heavier scalar state $\H$ or the pseudoscalar state $\A$,
whereas we impose the lightest $CP$-even state $\h$ to be the SM Higgs boson of
mass $m_{\h} = 125$~GeV. While the model features in general many free
parameters, they can all be reduced, for our purposes, to the value of the mixing
angle $\alpha$ and $\tan\beta$.

The way in which the mixing angles enter the couplings of the Higgs bosons to
the SM particles depends on the 2HDM configuration under consideration. For the
sake of the example, we consider in this section the $CP$-conserving version of
the Type~II 2HDM, where the
first Higgs field $\phi_1$ couples to the down-type quarks and the charged
leptons, and the second Higgs field $\phi_2$ couples to the up-type quarks, as
in the MSSM..

Whereas two different Higgs cascades can in principle be considered,
\be
  p p \to \H \to \A Z\to \h ZZ
  \qquad\text{and}\qquad
  p p \to \A \to \H Z\to \h ZZ \ ,
\ee
the absence of a $\H\h Z$ coupling in the Type~II 2HDM implies that the second
of the above processes is forbidden. In the notation of
Section~\ref{sec:indep}, this thus means that $H_1\equiv \A$ and $H_2\equiv \H$.
The corresponding production cross section
depends on the $\alpha$ and $\beta$ angle through the off-diagonal coupling
strengths of the Higgs bosons to the $Z$-boson $g_{\H\A Z}$ and $g_{\A\h Z}$,
\be
 g_{\H\A Z}=-\frac{g\sba}{2\cos\theta_w} \qquad\text{and}\qquad
 g_{\A\h Z}=\frac{g\cba}{2\cos\theta_w}\ ,
\label{eq:haz}\ee
with $g$ being the weak coupling and $\theta_w$ the electroweak mixing angle and the coupling of the Higgs bosons to $t\bar{t}$ and $b\bar{b}$ - see Sec.~\ref{sec:2hdm_results}.
While other Higgs production process could be relevant as
potentially yielding an observable signal (like the vector-boson fusion
production of an $\A$ boson), we opt to ignore them all as they would require
dedicated analyses which goes beyond the scope of this work.

\subsection{Higgs-Boson Production Cross Sections and Branching Ratios}
\label{sec:2hdm_results}
In order to evaluate the constraints that could be imposed on the 2HDM parameter
space from $\H$ cascades, we first need to calculate the $pp\to\H$ cross
section. We make use of the SM results~\cite{Dittmaier:2011ti,deFlorian:2016spz}
that we rescale by an appropriate loop factor,
\be
 \sigma(pp \to \H) = {\sigma_{\rm SM}}\times
   \frac{\left\vert \frac{\sin\alpha}{\sin\beta} F^{h}_{1/2}(\tau_t) + 
   \frac{\cos\alpha}{\cos\beta} F^{h}_{1/2}(\tau_b)\right\vert^2}
   {|F^{h}_{1/2}(\tau_t)+F^{h}_{1/2}(\tau_b)|^2},
\label{eq:Hsigma} \ee
where $\tau_f=4m_f^2/m_{\H}^2$ (with $f=t$, $b$) and where the loop function
$F^{h}_{1/2}$ is given by
\be
  \renewcommand{\arraystretch}{1.4}
  F^h_{1/2} = -2\tau \Big[1+(1-\tau) f(\tau)\Big] \qquad\text{with}\qquad
  f(\tau) = \left\{ \begin{array}{lc} 
    \left[ \sin^{-1} (1/\sqrt{\tau}) \right]^2 \ \ & \tau \geq 1\ , \\
     -\frac14 \left[ \ln\frac{1 + \sqrt{1 - \tau}}{1 - \sqrt{1 - \tau}}
       - i \pi \right]^2 \ \ & \tau < 1\ .
   \end{array} \right.
\ee

\begin{figure}
 \centering
  \includegraphics[width=0.48\columnwidth]{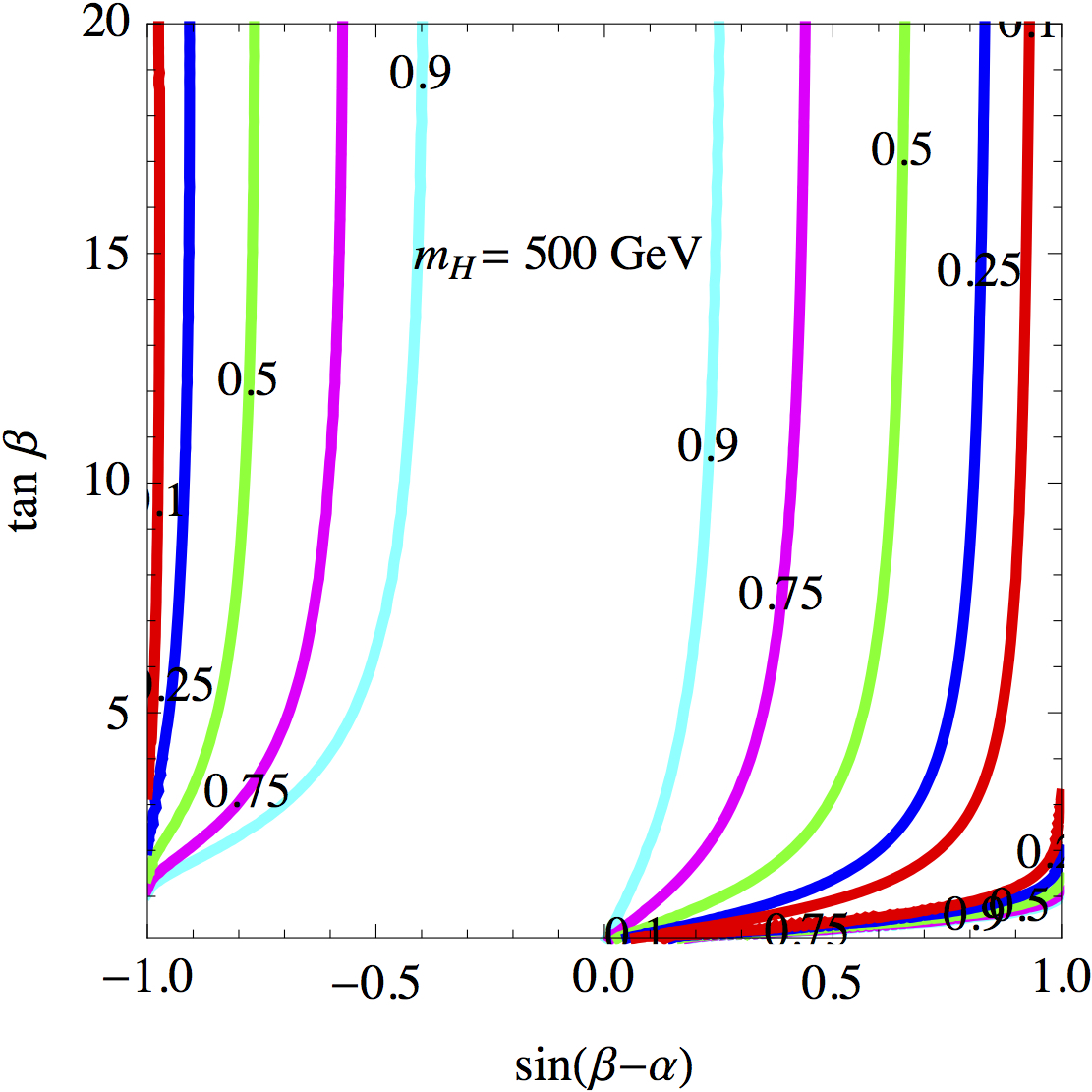}
  \includegraphics[width=0.48\columnwidth]{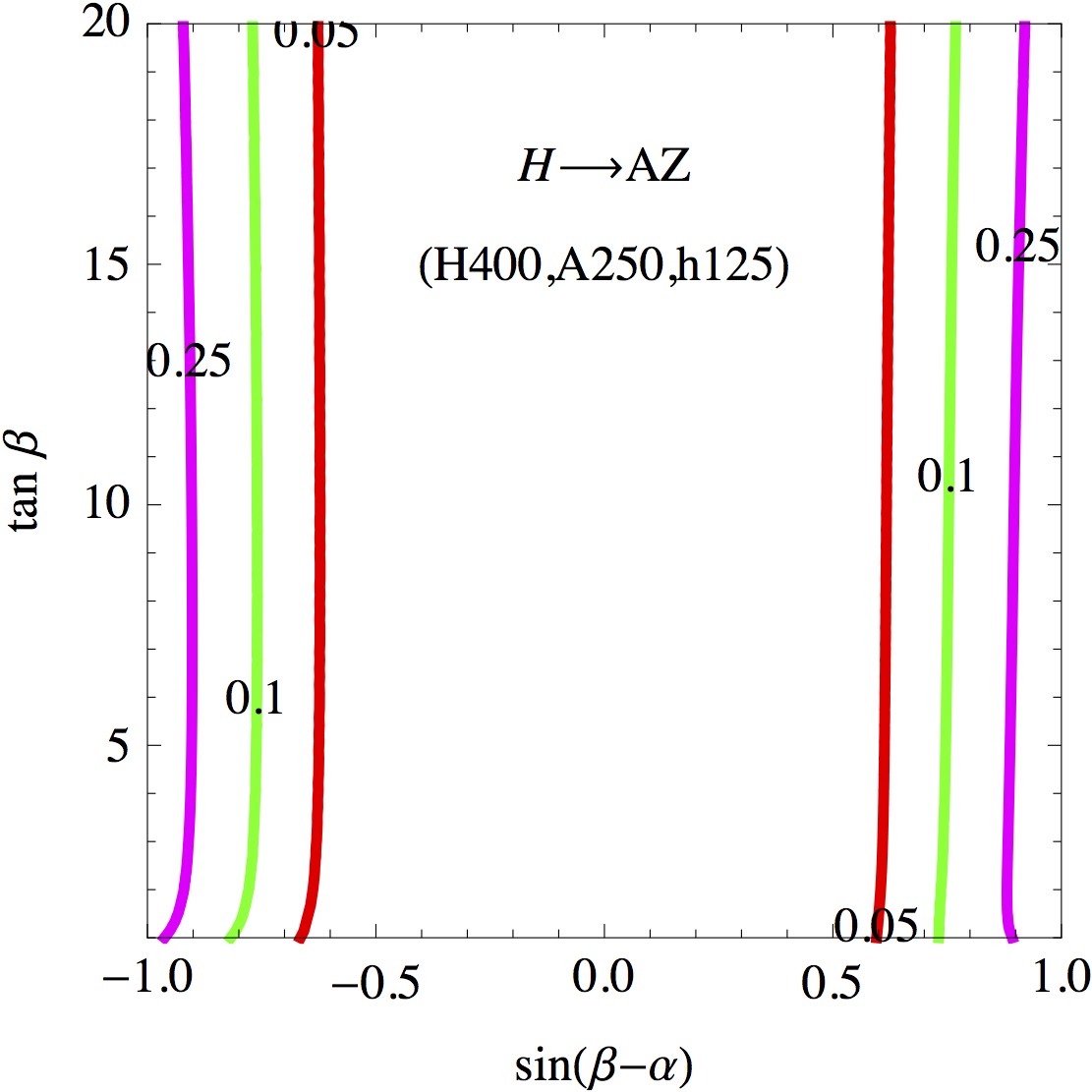}\\
  \includegraphics[width=0.48\columnwidth]{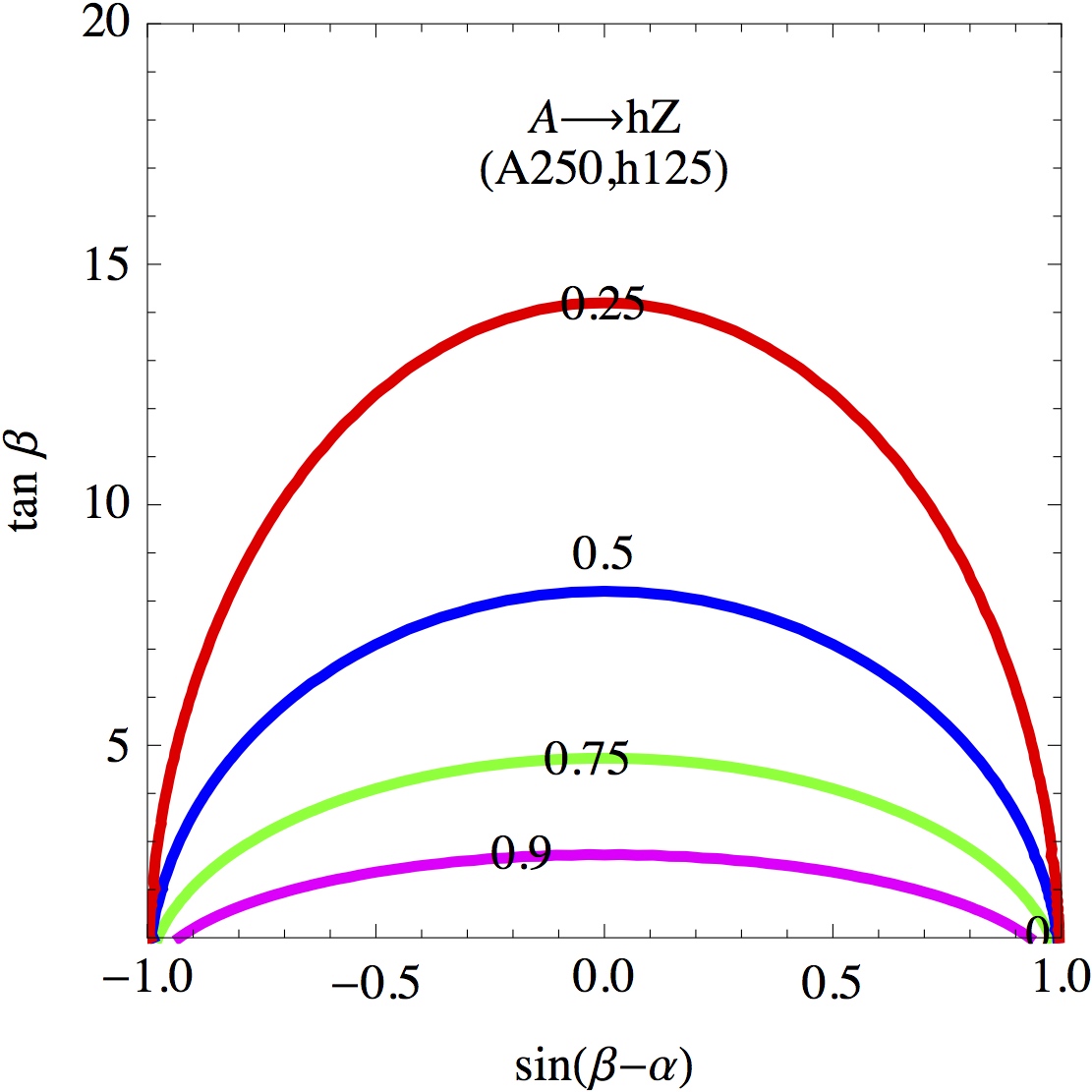}
  \includegraphics[width=0.48\columnwidth]{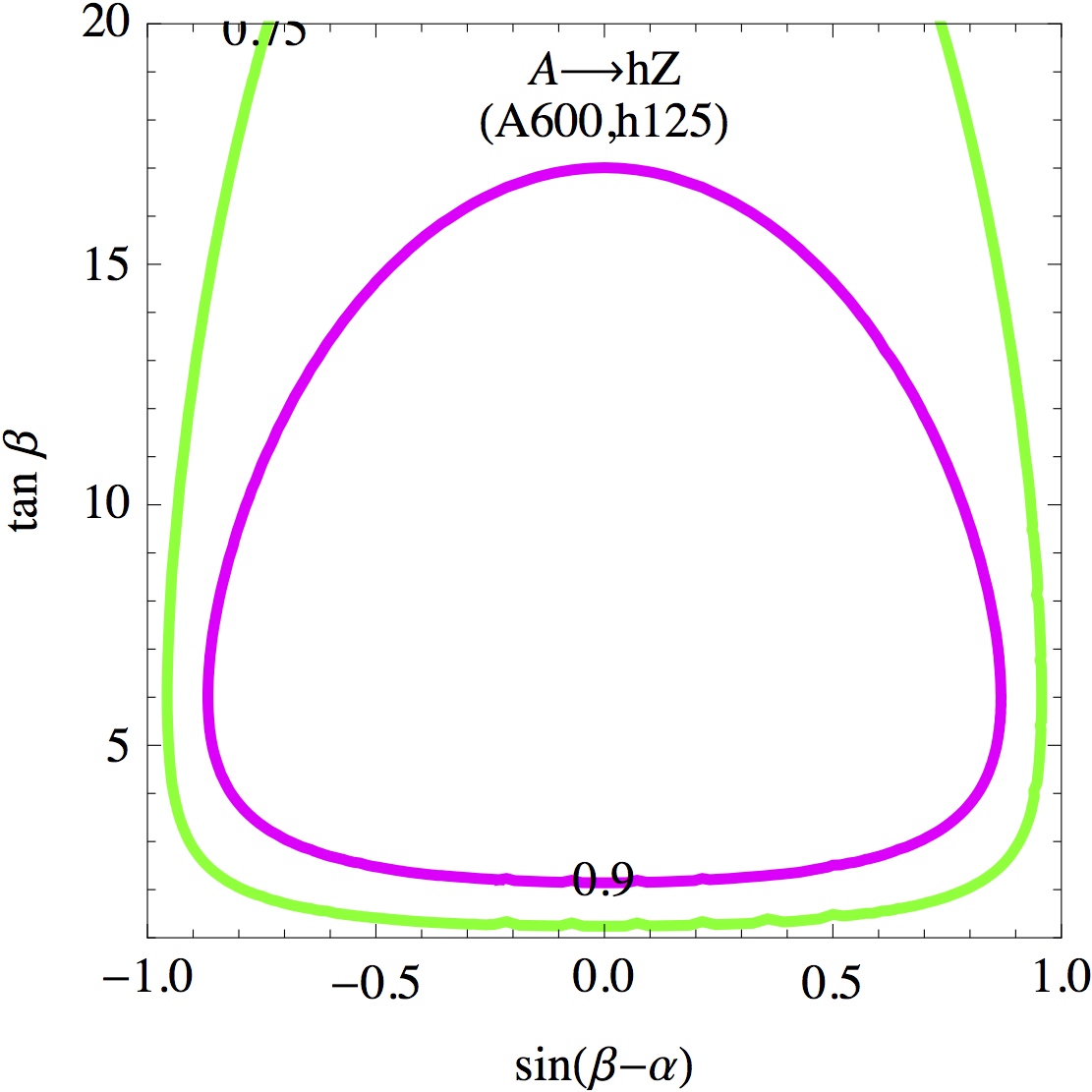}
  \caption{Dependence of the $\sigma(pp\to \H)$ cross section (upper left panel)
    and the $\H\to\A Z$ (upper right panel) and $\A\to\h Z$ (lower panel)
    branching ratios on the Higgs mixing angles $\alpha$ and $\beta$. The
    results are shown in the $(\sba, \tan\beta)$ plane and for the Higgs boson
    masses introduced in Section~\ref{sec:indep}. The cross section values
    (in the upper left figure) are
    normalized to the corresponding SM value for a SM Higgs-boson of 500~GeV.}
 \label{fig:SigmaH}
\end{figure}

In Figure~\ref{fig:SigmaH} (upper-left panel), we present, in a convenient
$(\sba,\tan\beta)$ plane, the dependence on the $\H$ gluon fusion production
cross section on the mixing angles for a heavy Higgs-boson mass of 500 GeV. The
results are normalized to the corresponding SM Higgs-boson production cross
section, and we observe that the cross section is maximum when $\sba\to0$ and
tends to vanish for $\sba\to \pm 1$. As the lightest Higgs boson $\h$ has to be
SM-like, $\sba\sim \pm 1$, some slight deviations being however still allowed by
current measurements~\cite{Coleppa:2013dya}. This constraint will nevertheless
be omitted from our analysis in which we aim to determine the constraints on the
parameter space that are issued solely from Higgs cascades at the LHC. The
asymmetry of the cross section dependence on $\sba$ (relatively to $\sba=0$)
originates from the $\alpha$ and $\beta$ dependence in
Eq.~\eqref{eq:Hsigma}. The cross section is hence enhanced both for small values
of $\tan\beta$ (due to an enhancement of the contributions of the top-quark
loops) and large values of $\tan\beta$ (due to an enhancement of the
contributions of the bottom-quark loops). The top-loop enhancement is more
pronounced in the positive $\sba$ half-plane, while the bottom-loop one impacts
the negative $\sba$ half-plane. Moreover, any further increase of $\tan\beta$
beyond 20 does not lead to any appreciable effect via the bottom loops, so that
we impose $\tan\beta<20$ in the following analysis.

The partial widths associated with the $\H\to \A Z$ and $\A\to \h Z$ decays are
controlled by the scaling factors $\sba$ and $\cba$ respectively, as illustrated
by Eq.~\eqref{eq:haz}. As a
result, the intermediate region in which $\sba$ is different both from 0
and $\pm1$ features an interesting enhancement of the $\H\to\h ZZ$ decay. In the
upper right and lower panels of Figure~\ref{fig:SigmaH}, we present contours of
specific branching ratios values for the $\H\to \A Z$ and $\A\to \h Z$ decays
for representative Higgs-boson mass choices corresponding to the
benchmark points introduced in Section~\ref{sec:indep}. As expected, we observe
that the $\H\to \A Z$ decay becomes prominent for $\sba\sim\pm1$, while the
$\A\to \h Z$ one exhibits a complementary behavior and becomes smaller in this
region. The qualitative difference in the behavior of the pseudoscalar decay
into a $\h Z$ pair for the {\bf BP1}-like (lower left panel) and {\bf BP2}-like
(lower right panel) configurations stems from the $t\bar t$ channel that is
kinematically open in the {\bf BP2} case and is dominant for low values of
$\tan\beta$. On the other hand, the decays into $b\bar{b}$ and $\tau^+\tau^-$
systems are enhanced for larger $\tan\beta$ values, and the partial width of
the $\A\to \h Z$ decay increases for $\sba\to0$, as shown in Eq.~\eqref{eq:haz}.
Consequently, the branching ratio associated with the $\A\to \h Z$ decay is
bounded from above in the large $\tan\beta$ region for both scenarios, as well
as for small $\tan\beta$ values in the {\bf BP2} case. This explains the origins
of the closed contours of given branching ratio values obtained for the
{\bf BP2} scenario. On different grounds, we have found that there is no
qualitative differences across scenarios for the $\H\to \A Z$ branching ratio.

\begin{figure}
 \centering
  \includegraphics[width=0.48\columnwidth]{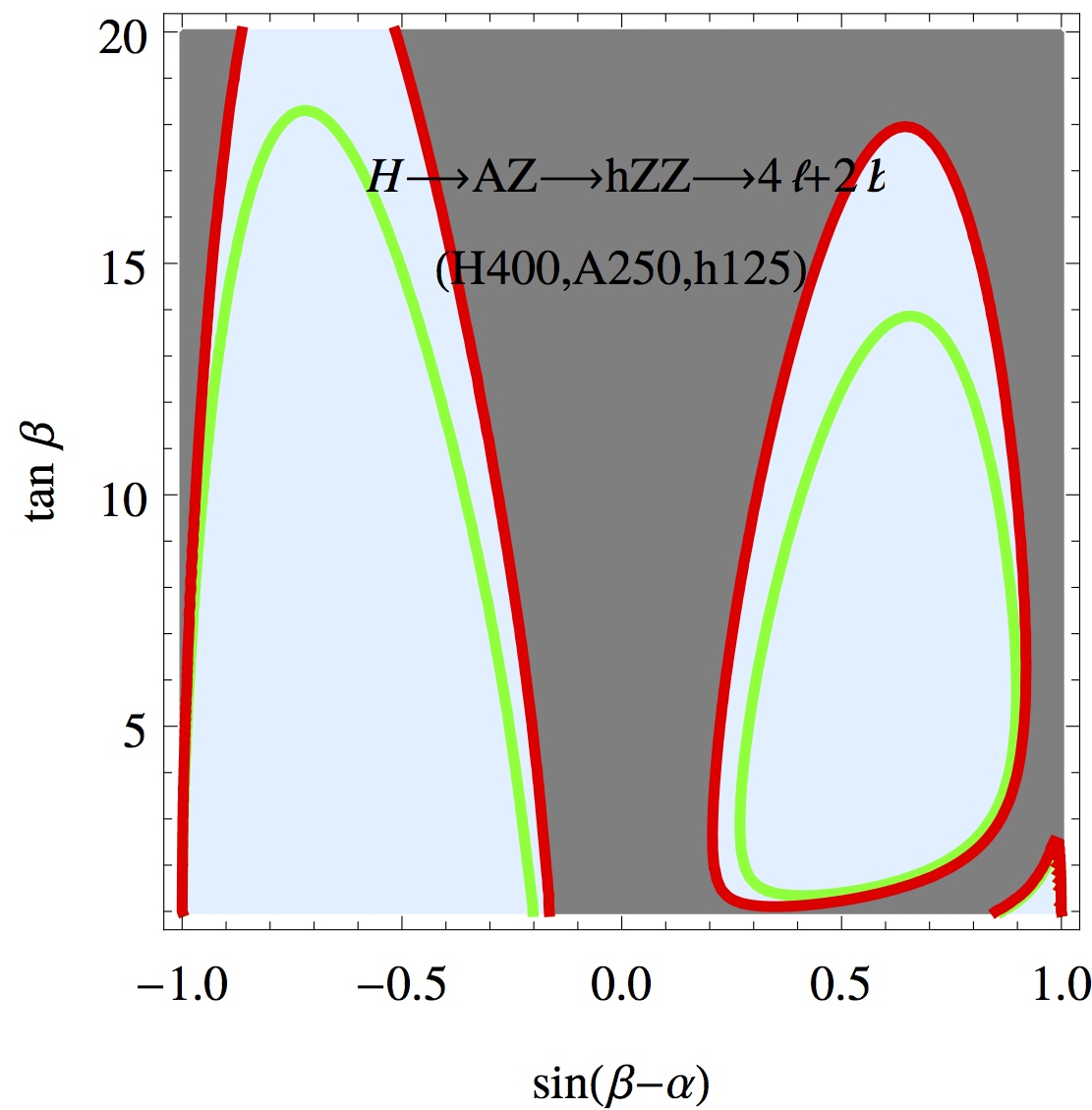}
  \includegraphics[width=0.48\columnwidth]{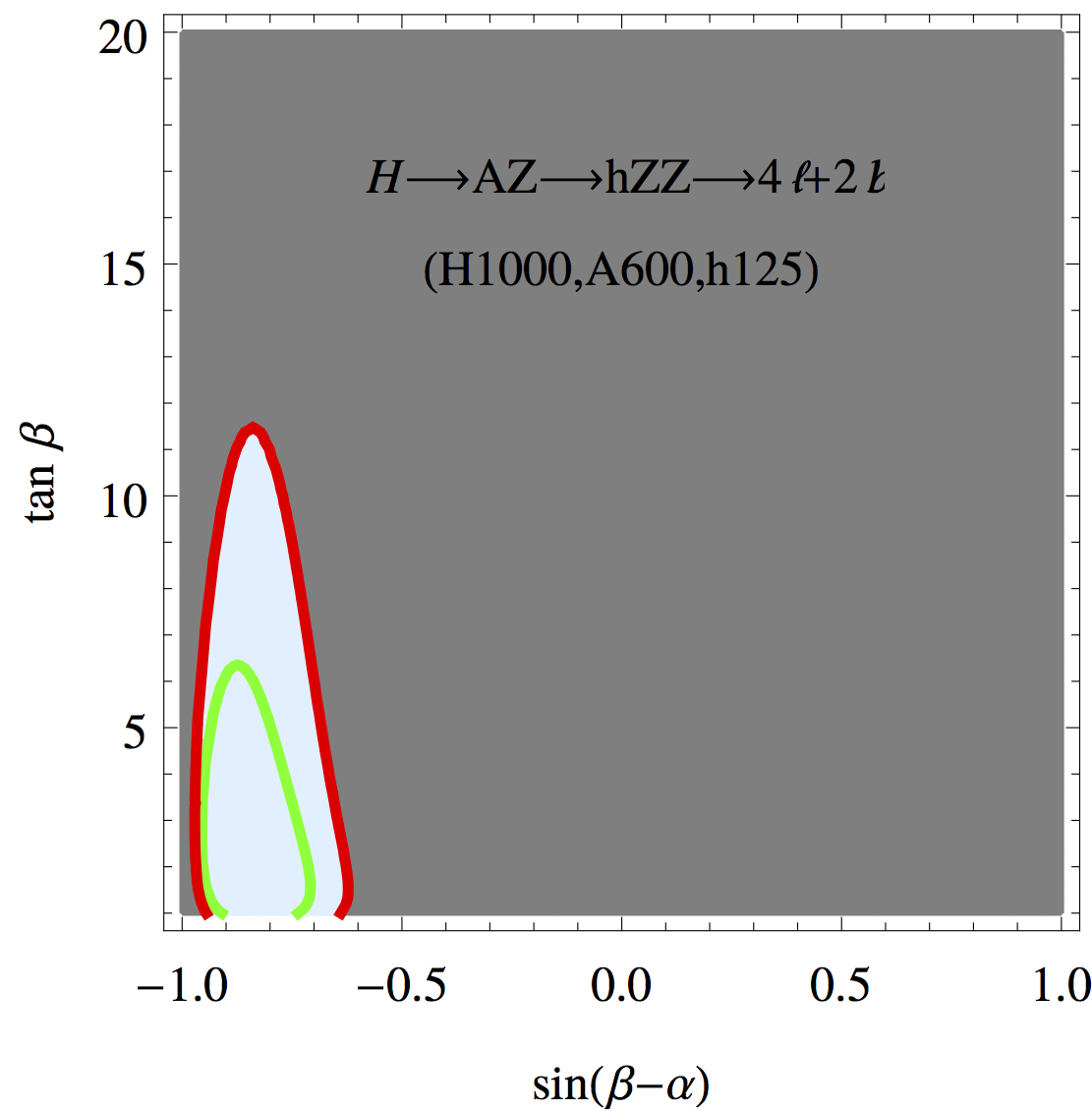}\\
  \includegraphics[width=0.48\columnwidth]{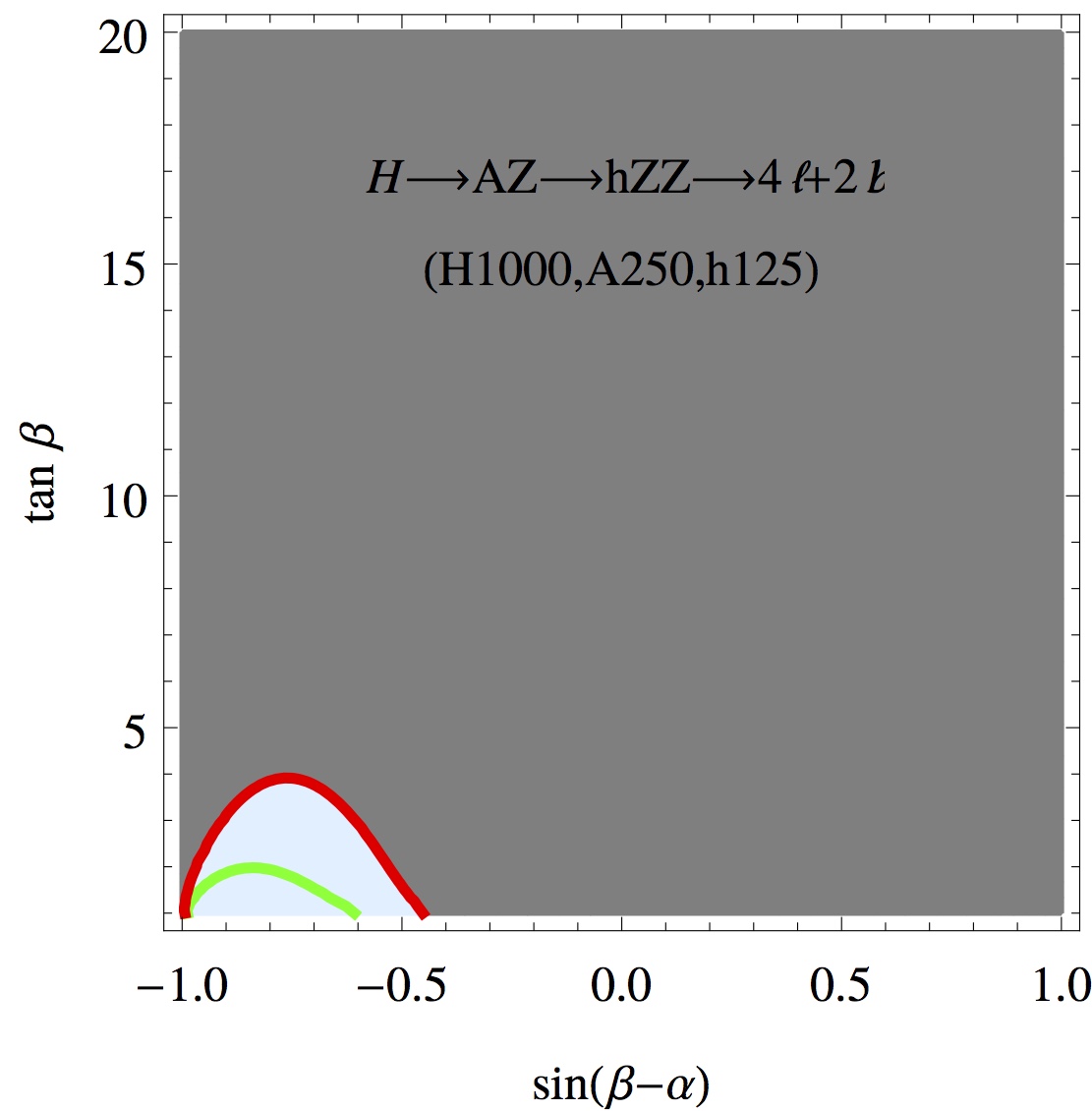}
  \includegraphics[width=0.48\columnwidth]{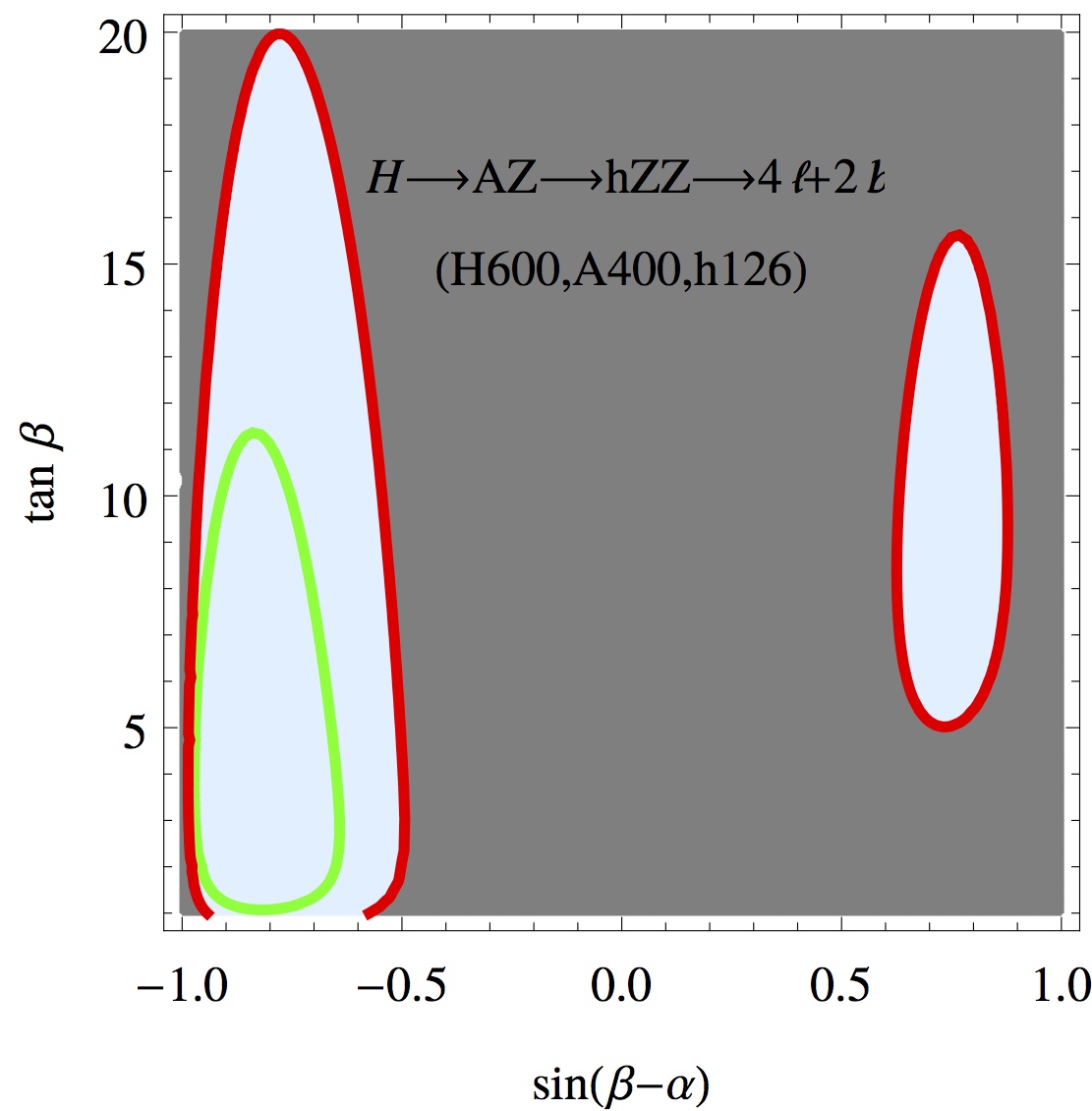}
 \caption{Sensitivity of the LHC in the 2HMD parameter space once all Higgs
   masses have been fixed as in the {\bf BP1} (upper left), {\bf BP2} (upper
   right), {\bf BP3} (lower left) and {\bf BP4} (lower right) scenarios. The
   green and red contours respectively correspond to the region covered by the
   $N(b)=2$ and $N(b)=1$ signal regions for an integrated luminosity of
   1000~fb$^{-1}$.}
 \label{fig:limit}
\end{figure}

In Figure~\ref{fig:limit}, we show the allowed regions in the $(\sba,\tan\beta)$
plane for mass configurations equal to those of the benchmark scenarios
introduced in the former section. If the final-state topology
is similar to the one encountered in the case of the considered benchmarks, the
already-computed upper limits on the signal cross section could be
applied. On the other hand, new limits could also be obtained after deriving the
selection efficiency that would be associated with the new signal, following the
different analysis strategies introduced in Section~\ref{sec:indep}.

The light blue regions shown on Figure~\ref{fig:limit} correspond to parameter
space configurations in which the cross section associated with the heavy
Higgs-boson cascade process is large enough to yield a $5\sigma$ discovery. The
results are based on the numbers quoted in the previous section and are
related to an integrated luminosity of 1000 fb$^{-1}$. The green contours
determine the reach of the $N(b)=2$ signal region of the $4\ell2b$ analysis
while the red ones refer to the $N(b)=1$ signal region of the same analysis. As
mentioned previously, the $4b2\ell$ analysis is not considered as it is expected to
lead to weaker bounds.

By virtue of a larger heavy-Higgs production cross section, the {\bf BP1} and
{\bf BP4} scenarios are much better covered, the $\H$ boson being indeed lighter
than
in the other cases. Moreover, while the functional form of the production cross
section favors the $\sba\approx 0$ region, the product of the two branching
ratio and their dependence on $\sba$ and $\cba$ moves the parameter space region
of interest away from the $\sba\sim 0$ region. The bulk of the discovery reach
is located, for the four benchmark, close to $\sba\sim 1$ that is precisely
the region favored by current Higgs data. On the other hand, the dependence on
$\tan\beta$ directly originates from the branching ratio results of
Figure~\ref{fig:SigmaH} that show that large and small $\tan\beta$ values may
respectively imply a reduced sensitivity due to the importance of the $\A\to
b\bar{b}$ decay and $\A\to t\bar t$ decay (if relevant). While the rates of the
cascade-decay processes undergone by heavier Higgs bosons may be suppressed,
it is seen that they are indeed a viable option to find these additional Higgs bosons at the
LHC particularly if they are moderately heavy. The expected suppression of the
branching ratio has indeed been found not to be sufficient to balance the strength
of simple selection cuts allowing for the separation of the signal from the
background.

\section{Conclusions}
\label{sec:conclusions}
While the spectrum of the Standard Model has been established firmly today,
physics beyond the Standard Model still remains a mystery. On the theoretical
side, creative model building has explored avenues with an enlarged gauge group,
extended matter representations and often a richer Higgs sector. While dedicated
analyses are necessary to probe specific models of new physics, many models
share common features (at least in terms of their spectra) so that they could be
explored simultaneously in a general manner. In this spirit, this paper aims to
study heavy neutral Higgs bosons that cascade decay into SM particles via
intermediate lighter scalar states,
as could occur in varied new physics theories. While exotic Higgs-boson decays
have been investigated in the literature, doubly-exotic modes involving
several Higgs bosons have mostly not been targeted widely so far although they are an interesting probe for potential discovery.

In this work, we have discussed the generic cascade decay process
$pp\to H_2\to H_1 Z\to \h ZZ$ where a heavy Higgs boson $H_2$ decays into a
lighter Higgs boson $H_1$ and a $Z$-boson, and where the $H_1$ boson further
decays into a SM Higgs-boson $\h$ and a $Z$-boson. Investigating a final-state
signature made of either two $b$-jets and four charged leptons, or of four
$b$-jets and two charged leptons, we have found that the discovery potential of
such a process heavily depends on the magnitude of the mass splittings between
the different scalar states, which directly impacts final-state object
identification. It turned out that the $2b4\ell$ channel is very promising, in
particular when the requirement on the number of $b$-tagged jets is relaxed to
$N(b)=1$. Although this channel does however not allow for the proper
reconstruction of the heavy Higgs bosons, it provides an excellent handle for
exhibiting the presence of a new physics signal. In contrast, the $4b+2\ell$
final state turns to be less promising,
due to the non-perfect $b$-jet identification and the larger backgrounds.

We have begun with performing our collider analysis in a simplified-model
approach inspired by the 2HDM, without resorting to specific values for the new
physics couplings. This has allowed us to design several dedicated analyses,
optimizing them for a good Higgs-cascade signal selection efficiency and an
important associated background rejection. We have then applied our findings
to assess the LHC discovery potential of a specific model that has been taken
for the sake of the example to be the Type-II 2HDM. In this theoretical
framework,
we have found that the LHC is sensitive to Higgs-to-Higgs cascades in particular
if the heaviest scalar state mass is moderate and for couplings close to those
currently allowed by LHC Higgs data. This preferred configuration enhances on
the one hand the heavy Higgs boson production cross section, and guarantees on
the other hand that the decay products of the Higgs boson can properly be
reconstructed.
2HDM compressed scenarios like our {\bf BP1} benchmark point satisfy both these
criteria and are understandably expected to be better covered by future LHC
results. Scenarios with a slightly heavier spectrum but exhibiting not too
large mass splittings, like
our {\bf BP4} scenario, are expected to be well probed too, however with a 
more limited reach. Finally, the sensitivity to scenarios like our {\bf BP2} and
{\bf BP3} where the spectrum is much heavier (the heaviest state being at
the TeV scale) is still appreciable but reduced as a consequence of the Higgs
decay products being in a boosted regime for which our analysis is not sensitive
to and the smaller production cross section.

Higgs cascades therefore offer a new channel to look for extended scalar
sectors, complementing and potentially competing - at least in some models where
heavier Higgs bosons for instance feature reduced couplings to fermions - with
the more traditional approaches seeking heavier Higgs bosons.

\begin{acknowledgments}
SS would like to thank J.~Beuria and P.~Sahoo for help with {\sc Madgraph} and
ROOT. BC acknowledges support from the Department of Science and Technology, India,
under Grant YSS/2015/001771. The work of BF has been partly supported by French
state funds managed by the Agence Nationale de la Recherche (ANR), in the
context of the LABEX ILP (ANR-11-IDEX-0004-02, ANR-10-LABX-63).
\end{acknowledgments}

\bibliography{biblio}

\end{document}